\documentclass[fp,twocolumn]{jpsj3}
\usepackage{graphicx}
\usepackage{bm}

\newcommand{\simj}{\stackrel{>}{_\sim}}
\newcommand{\simk}{\stackrel{<}{_\sim}}

\begin{document}
\title{ Melting Points and Formation Free Energies of Carbon Compounds with Sodalite Structure}

\author{Kazuhiro Sano and Kenshin Nato}
\inst{Department of Physics Engineering, Mie University, Tsu, Mie 514-8507, Japan}

\abst{
\quad Using  first-principles calculations, we  investigated the melting temperatures  $T_{\rm m}$ and formation free energy of carbon compounds  with sodalite structures, $X$C$ _6$, $X$C$ _{10}$,  and $X$C$ _{12}$, where $X$ is F, Na, Cl, or so on. 
These compounds are expected to be phonon-mediated superconductors exhibiting high transition temperatures $T_{\rm c}$ of up to about 100 K.
We estimated  $T_{\rm m}$  as a function  of  pressure $P$ by  the first-principles molecular dynamics method  and showed the results as phase diagrams   on the $P$-$T$ plane  together with the results of  $T_{\rm c}$.
The phase diagrams  indicate  that the $T_{\rm m}$ of NaC$_{\rm 6}$, which has a $T_{\rm c}$ of  about  100 K,    is about $1300$ K or more at $P=30$ GPa. Furthermore, the $T_{\rm m}$ of FC$_{\rm 6}$  is  about 2200 K even at $P=0$ GPa, where its $T_{\rm c}$ is about 80 K. 
Similar results were obtained for  FC$_{\rm 10}$ and ClC$_{\rm 10}$ systems.
These results suggest that some compounds can  stably exist as high-temperature superconductors even at room temperature and pressure.
To examine the feasibility of synthesizing these compounds, we  estimated the  formation  enthalpies and     formation  free energies.
The results suggest that NaC$_6$ can be formed under a sufficiently  high pressure of about  300 GPa and a high temperature of about 6500 K.
}

\maketitle
\section{Introduction}\label{intro}
Since the high-temperature superconductors of hydrogen compounds with a sodalite structure, such as  YH$_6$ and lanthanum decahydride (LaH$_{10}$),\cite{Wang-2012,Feng-2015, Y-Li-2015, H-Liu-2017, Peng-2017, Heil-2019,Kruglov-2019-LaH10} were  predicted to indicate  $T_{\rm c}$  over 250 K,  much effort has been made to clarify  the superconductivity of these  compounds and/or to find new  hydrogen compounds.\cite{Wang-2012, Feng-2015, Y-Li-2015, H-Liu-2017, Peng-2017, Heil-2019,Kruglov-2019-LaH10}.
Based on  theoretical results\cite{Wang-2012,Feng-2015, Y-Li-2015, H-Liu-2017, Peng-2017}, experiments\cite{Drozdov-LaH10,Somayazulu-2019,Troyan-2019-YH6,Kong-2019-YH6} succeeded  in finding   these superconductors whose  $T_{\rm c}$ is almost close to the result of  first-principles calculations.
This high-temperature superconductivity (HTS)   is  caused by  phonon-mediated attraction, and the mechanism of the superconductivity is conventional. 
Therefore, $T_{\rm c}$ is determined mainly by two parameters: the electron-phonon coupling constant $\lambda$ and the characteristic phonon frequency $\omega_{\rm log}$.
 When both parameters are simultaneously  large, HTS can be expected. 
In fact,  many hydrogen compounds with a sodalite structure have  a large electron-phonon coupling constant $\lambda$, that is larger than about 2.0.\cite{Wang-2012,Feng-2015, Y-Li-2015, H-Liu-2017, Peng-2017, Heil-2019}
Typical values of $\omega_{\rm log}$ are around 1000 K.

In  these hydrogen compounds,  high pressure may stabilize these characteristic structures and lead to a high phonon frequency.
These materials are only stable over a limited pressure  and  the phonon frequency decreases with pressure.\cite{Wang-2012,Feng-2015, Y-Li-2015, H-Liu-2017, Peng-2017, Heil-2019}
Therefore, hydrogen compound HTS can only be achieved with devices using diamond anvil cells at this stage.
To realize practical devices, the required pressure should be as low as possible, preferably  normal pressure.

A material constructed with carbon atoms, such as diamond,  has a high phonon frequency  up to about 2000 K  at atmospheric pressure.
Boron-doped diamond  has been  studied  as  a candidate of   phonon-mediated HTS, and its   $T_ {\rm c} $ reaches  $\sim 25$ K at atmospheric pressure.\cite{Ekimov2004,Ma,Kawano,Okazaki}
Intercalated graphite  and alkali-doped C$_{60}$ compounds are  known carbonate  superconductors with high $T_ {\rm c} $ of up to $\sim 33$ K.\cite{Weller,Smith,Hao2023a,Ramirez}

 On the basis of the intuitive idea that the sodalite structure plays an important role in HTS,
 carbon compounds with a sodalite structure have been proposed as candidates for HTS materials.\cite{Lu-2016,Wei-2016,Sano2022,Khan2022,Hao2023b,Jin2024}
Based on the density functional theory, materials  composed only  of carbon atoms are known to be insulators with a large charge gap.\cite{Pokropivny}
By combining them with another element, $X$,  to form a compound, carriers are introduced into the system\cite{Sano2022,Khan2022,Hao2023b,Jin2024}, where $X$ is Li, Na, Cl, or so on.
The crystal structures of the compounds are shown in Fig. 1, where (a) $X$C$_{6}$ and (b) $X$C$_{10}$ are shown as representative examples.
 $X$C$_{12}$ has a structure similar to that of $X$C$_{6}$    shown in Fig. 1(a), except for one $X$ atom.

%
\begin{figure}[hb]
\begin{center}
\includegraphics[width=0.8 \linewidth]{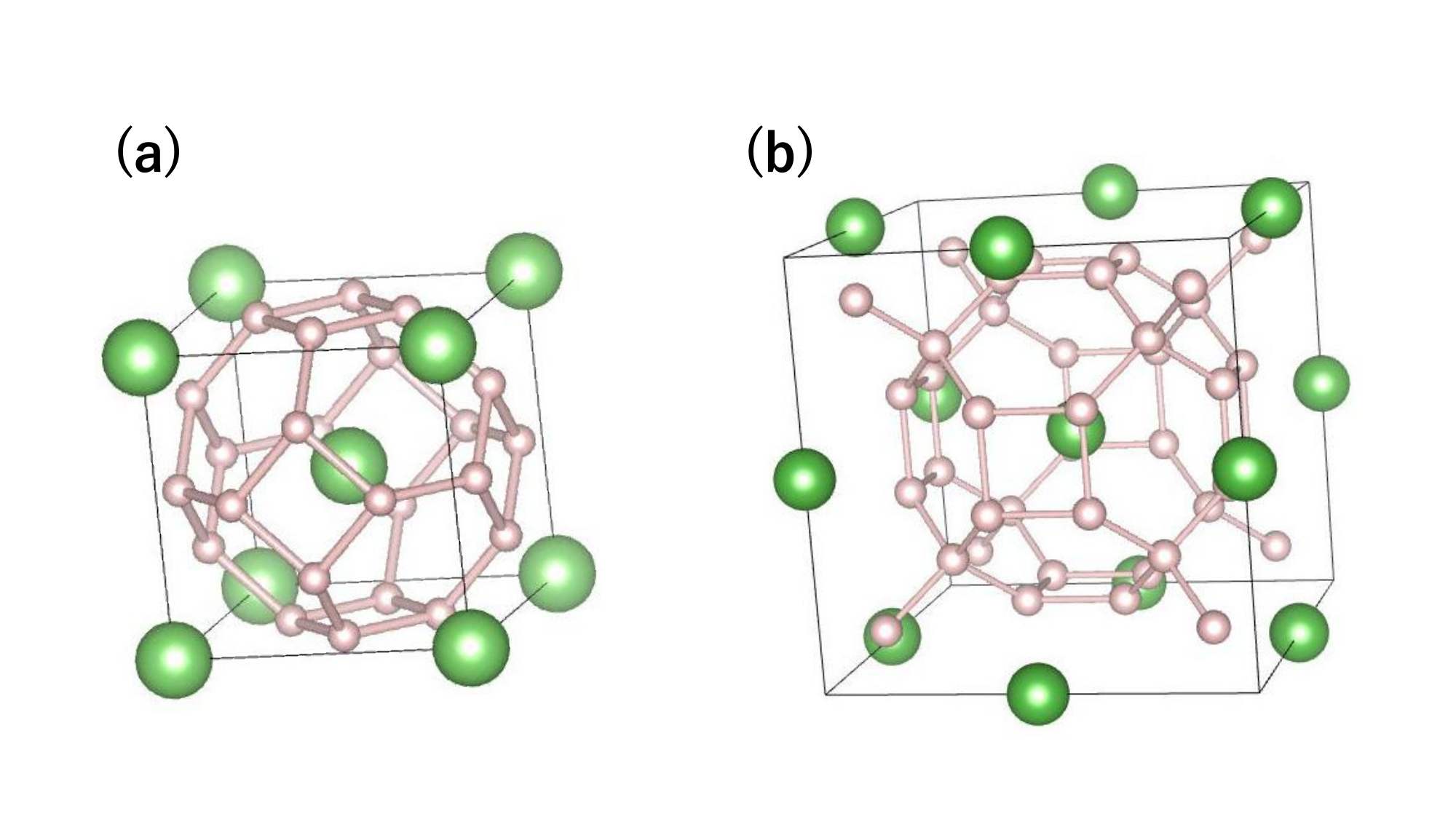}
\end{center}
\caption{(Color online)
   Structures of  the sodalite-type compounds (a) $X$C$_6$  and  (b) $X$C$_{10}$, where large spheres represent $X$-atoms and small spheres are carbon.  Here, the crystal structure of $X$C$_{12}$ corresponds to (a) with one $X$ atom removed.
}
\label{model}
\end{figure}
By first-principles calculations, it has been shown that some of these compounds exhibit $T{\rm c}$  of about 100 K even at atmospheric pressure.\cite{Lu-2016,Sano2022,Khan2022,Hao2023b,Jin2024}
However, the existence of $X$C$ _6$, $X$C$ _{10}$, and $X$C$ _{12}$    has  not yet been experimentally demonstrated. 
Furthermore, their stabilities at finite temperatures neither have been  clarified.
Even if these are stable near $T=0$, they will not be practical if the crystals become unstable below room temperature.

 In this study,  we investigated  the stability of  carbon compounds with  sodalite structure  by the  first-principles molecular dynamics (FPMD)  method.
First, we estimated the temperature $T_{\rm u}$ at which the system becomes unstable.
Then, we  estimated $T_{\rm m}$ using the known relationship between $T_{\rm u}$ and the melting point $T_{\rm m}$.
These results are shown as  phase diagrams for  compounds  on the $P$-$T$ plane,   together with the superconducting transition temperature $T_{\rm c}$.

We also calculated the  formation  energies of   compounds from the ground state energies of the elements.
To consider the effect of  finite temperature, we obtained  the  free energies of  NaC$ _6$  and  FC$_{\rm 6}$ as  typical cases.
From these results, we estimated the Gibbs free energies and discussed whether the formation of these compounds is possible.
%
\section{Model and Methods}\label{model-section}
Calculations were performed using  Quantum ESPRESSO (QE), which is an integrated software of open-source computer codes for electronic-structure calculations.\cite{QE} 
In the FPMD calculations, we used the $N,P,T$=const. ensemble system, where the system is a supercell made up of multiple unit cells,  as shown in Fig. 1.
To reduce finite size effects, the direction vector that determines the shape of the supercell is set to change at each time step of the simulation.
This variable cell method includes system fluctuations better than the method using fixed direction vectors.

To estimate $T_{\rm m}$, we used a combination of the FPMD method and the Z-method \cite{Belonoshko2006}, which is  used to determine melting points.\cite{Finneya2011,  Belonoshko2012,Haskins2012, Wang2013,Geng2015, Anzellini2019}
Our method uses FPMD simulations to determine the critical temperature $T_{\rm u}$ at which the crystal becomes thermally unstable, and then calculates $T_{\rm m}$ from the relationship $T_{\rm m} \simeq T_{\rm u}/1.23$\cite{Belonoshko2006}.
Details of the Z-method are discussed in  Appendix A.

When the temperature of a system exceeds $T_{\rm u}$, the motion of atoms in the system changes significantly in a short period of time. Therefore,   it is easy to find $T_{\rm u}$ in the simulation.
 To obtain $T_{\rm u}$,  we  calculated the Lindemann  parameter defined as $L_p=\sqrt{\langle {\bm r}_i^2 \rangle}/d$.\cite{Lindemann1910,Ross1969,Shapiro1970}
Here, ${\bm r}_i$ is the displacement of each atom  from its equilibrium position at $T = 0$,  and $d$ is the distance between the nearest-neighbor atoms in the crystal at $T=0$.   The thermal average is taken over all atoms at each time step  in the simulation.

When the value of $L_p$ exceeds a certain critical value, the system  becomes  unstable rapidly\cite{Jin2001}.  Thus, we calculated $L_p$  as a function of temperature and found $T_{\rm u}$  from the point where it changes significantly.
%
\begin{figure}[hb]
\begin{center}
\includegraphics[width=0.8 \linewidth]{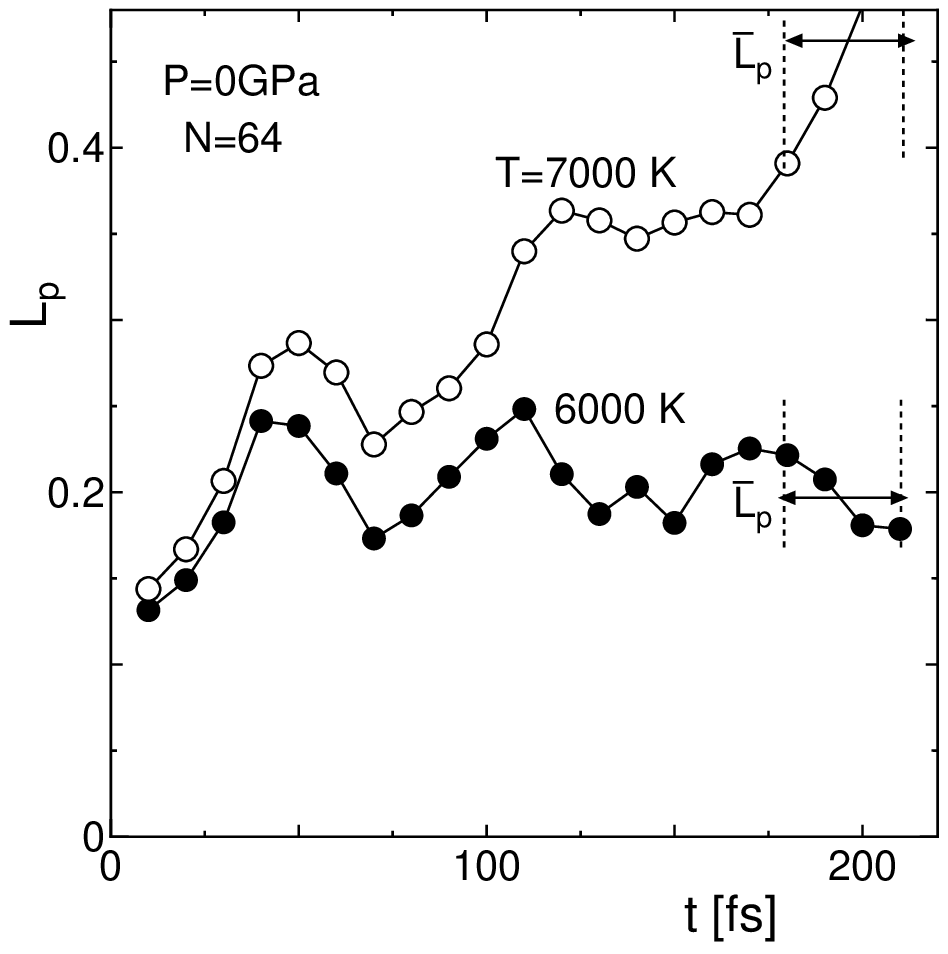}
\end{center}
\caption{ Lindemann  parameter $L_p$ as a function of  simulation time $t$, where solid circles and  empty  circles represent the results for $T=6000$  and  $7000$ K, respectively.  $\bar{L_p}$ represents the average of the last four points of $L_p$.
 }
\label{Lp-time}
\end{figure}
%
%
\begin{figure}[hb]
\begin{center}
\includegraphics[width=0.8 \linewidth]{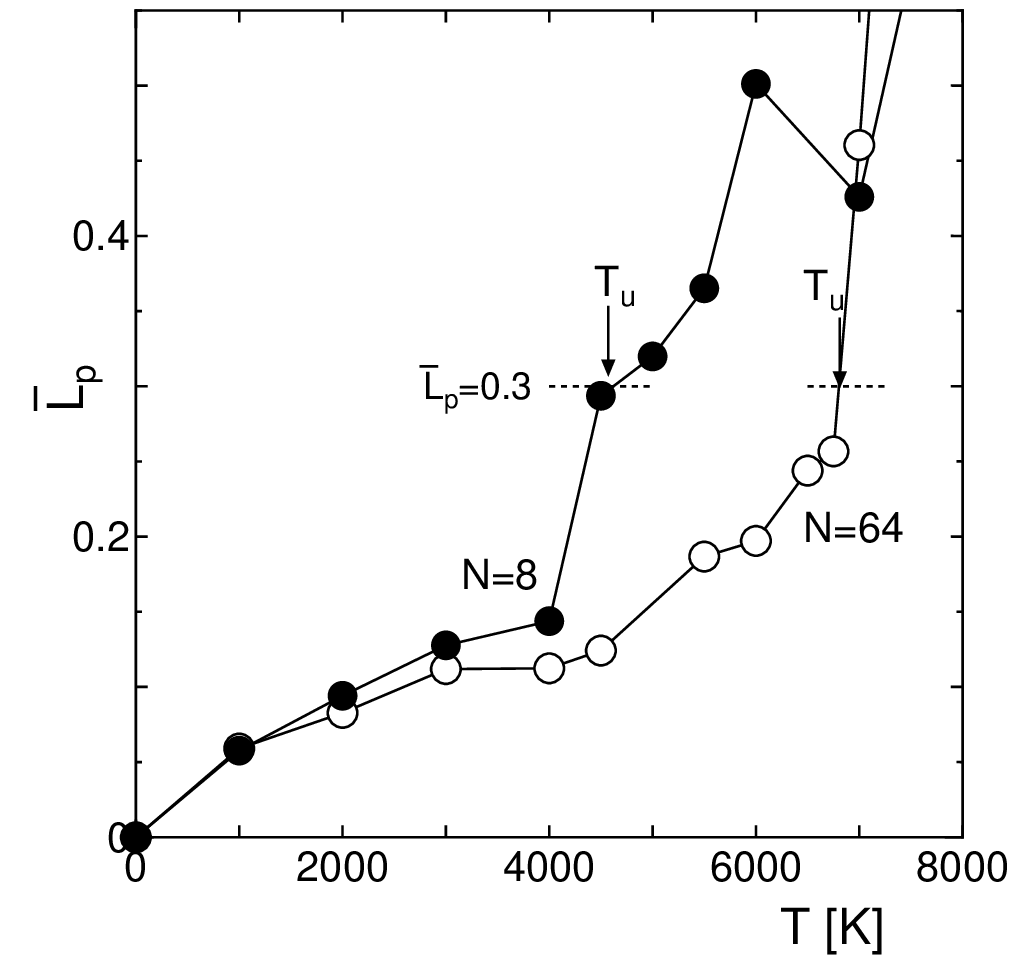}
\end{center}
\caption{ $\bar{L_p}$ as a function of $T$, where solid circles and  empty  circles represent the results for $N=8$ and  64, respectively.
$T_u$ is determined by  the condition $\bar{L_p}=0.3$. 
}
\label{Lpbar}
\end{figure}
%
%
\begin{figure}[th]
\begin{center}
\includegraphics[width=0.8 \linewidth]{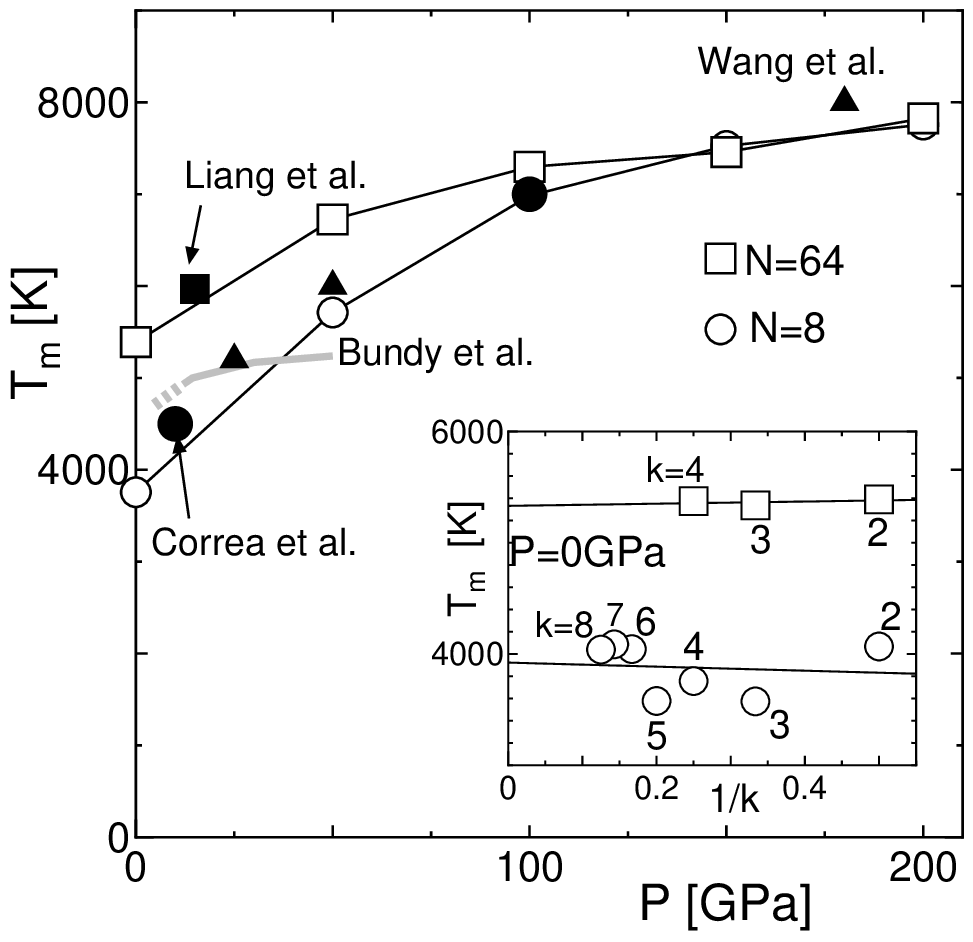} 
\end{center}
\caption{Melting points of diamond as a function of pressure $P$,  where empty circles and empty squares represent  $T_{\rm m}$ for the $N=8$ and 64 systems, respectively. Solid squares, solid circles, and solid triangles represent the results of  Liang et al. \cite{Liang2019}, Correa et al.\cite{Correa2006},  and Wang  et al, respectively\cite {Wang2005}.  The light and broken lines show the result of Bundy et al. \cite{Bundy1996}, where the  broken line indicates the phase boundary between graphite and diamond.
The inset shows the $k$-dependence of $T_{\rm m}$ at $P=0$ GPa, where $k$ is the mesh number in   wavenumber space.
}
\label{dia-souzu}
\end{figure}
%
%

At first, we considered the diamond system to test the above method.
Figure \ref{Lp-time} shows the Lindemann  parameter $L_p(t)$ as a function of  simulation time $t$ for  a diamond system with $N=64$ at $P=0$ GPa, where $N$ is the number of atoms in the system.  Here, solid circles and  empty  circles represent the results for $T=6000$ and  $7000$ K, respectively.
It indicates that the values of $L_p(t)$ at $T=6000$ K  are almost constant   with $t$,  and those  at $T=7000$ K increase with $t$ and seem to diverge for $t \simj 170 fs$.
This result suggests that $T_{\rm u}$ is between 6000  and 7000 K.

To analyze  $T_{\rm u}$ in  more detail, we defined  $\bar{L_p}$ as the average value  of  $L_p(t)$ at  the last four points ($ t =180,190,200,$ and 210 fs), as shown in Fig. 2.
As shown in the figure, it is expected that the system will reach a certain degree of equilibrium after about 150 fs. Therefore, we used $\bar{L_p}$ as the approximate equilibrium value of $L_p(t)$.

Figure 3 shows   $\bar{L_p}$ as a function of $T$ for the systems with $N=8$ (solid circles) and 64 (empty circles). 
The value of  $\bar{L_p}$ increases with temperature and appears to diverge above  $\bar{L_p} \sim 0.3$.\cite{Jin2001}
Hereafter, we assumed the temperature corresponding to $\bar{L_p}=0.3$ to be $T_{\rm u}$, where the point of $\bar{L_p}=0.3$ is determined by interpolation.
As shown in the figure, we found that $T_{\rm u} \simeq 4500$ K for $N=8$ and  6800 K for $N=64$. 
This leads to $T_{\rm m} \simeq 3700$ K for $N=8$ and  5500 K for $N=64$.
When the system is small,  fluctuations of the system are large and $T_{\rm u}$ tends to be low.

 Figure 4 shows the $T_{\rm m}$ of diamond as a function of pressure $P$ along with  previous theoretical\cite{Wang2005,Correa2006} and experimental\cite{Bundy1996,Liang2019} results.
Here,  the solid  and  empty  circles represent the results for $N=8$ and 64, respectively.
In the FPMD calculation, we used mesh numbers  $k=6$ and 2 in   wavenumber space   for the systems with $N=8$ and 64, respectively.
The obtained results  are almost in agreement with  those of  previous works.

The figure also shows that the size effect for $T_{\rm m}$ is greatest when $P=0$  and decreases as the pressure increases.
This is probably due to the decrease in the fluctuations of the system with increasing pressure.
The inset shows the $k$-dependence of  $T_{\rm m}$ for  the systems with $N=8$ and 64 at $P=0$.
This indicates that the $k$-dependence is rather small compared with the size effect in both systems.
%
%
%
\section{Results and Discussion }\label{result}
\subsection{$T_{\rm m}$ as a function of $P$  for  C$_6$ and C$_{10}$}\label{XC6}
We applied the above method to  pure carbon systems C$_6$ and C$_{10}$ as reference systems for $X$C$_6$ and $X$C$_{10}$.
 Figure 5(a)  shows the $T_{\rm m}$ of C$_6$ as a function of pressure $P$, where   empty  circles, empty squares, and solid triangles represent the results for $N=12$,  48, and 96, respectively.
The $T_{\rm m}$ of C$_6$ is at least 2000 K at P = 0 and  increases slightly with pressure.
Although its $T_{\rm m}$ is lower than that of diamond, it is stable at room temperature and pressure.

The figure indicates that  $T_{\rm m}$ for each size  is almost in agreement with each other  at all pressures, and the size dependence seems not to be  so large.
However,  the size effect of  $T_{\rm m}$ is not simple. At a certain pressure, $T_{\rm m}$ is high when the size is large. However, the opposite  also appears depending on the pressure. 
The use of variable  supercells is speculated to be the cause of the complex size effects.

Figure 5(b) shows the $T_{\rm m}$ of C$_{10}$ as a function of pressure $P$, where   solid circles   and solid squares represent the results for $N=10$  and 40, respectively.
 $T_{\rm m}$  is about  3500 K at P = 0 and  increases slightly with pressure  up to about 100 GPa, but decreases  above that.
This suggests that  C$_{10}$  would be able to exist stably at room temperature and pressure, although it has not yet been experimentally discovered.
  
%
\begin{figure}[h]
\begin{center}
\includegraphics[width=0.8 \linewidth]{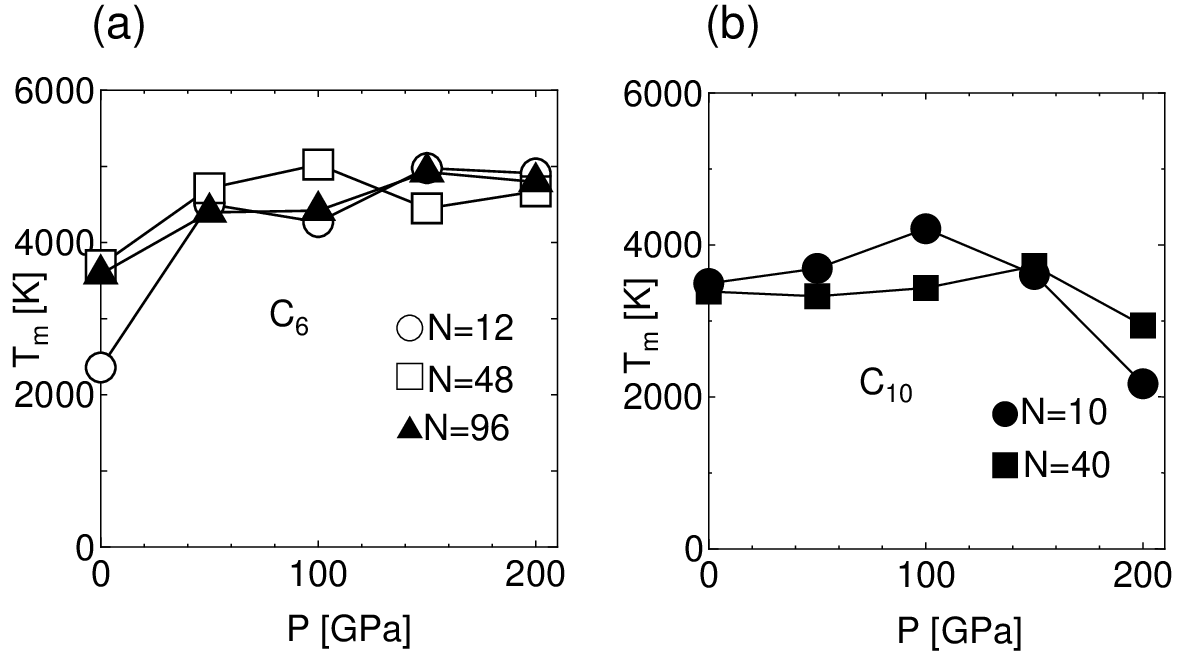}
\end{center}
\caption{(a) $T_{\rm m}$  of C$_6$ as a function of  $P$,  where empty circles, empty squares, and solid triangles  represent  $T_{\rm m}$ for the $N=12$,  48, and 96 systems, respectively.
(b)  $T_{\rm m}$ of C$_{10}$ as a function of  $P$, where  solid circles  and solid squares  represent  $T_{\rm m}$ for the $N=10$  and 40 systems, respectively.
  }
\label{C6-C10-souzu}
\end{figure}
%
%
\subsection{ $T_{\rm m}$  and $T_{\rm c}$ as  functions of $P$  for   $X$C$_6$  and   $X$C$_{10}$  }\label{Tc1}
We considered compounds of  $X$C$_6$  and   $X$C$_{10}$, combining carbon and $X$-atoms.
Mainly,  we presented the results for compounds  NaC$_6$, FC$_6$,  FC$_{10}$,  and ClC$_{10}$ as  typical cases.

Figure 6(a) shows the  $T_{\rm m}$ of NaC$_6$ as a function of pressure $P$, where solid circles and  solid squares represent the results for $N=14$ and 56, respectively.
In addition to $T_{\rm m}$, the superconducting transition temperature $T_{\rm c}$\cite{Allen1975} as a function of pressure $P$ is also shown as empty circles,  where  $T_{\rm c}$ is multiplied by 10 for ease of viewing.
%
Here, $T_{\rm c}$ is calculated by the tetrahedral method \cite{QE}, which allows us to avoid the ambiguity in the choice of Gaussian  broadening that exists in the interpolation method.

\begin{figure}[bh]
\begin{center}
\includegraphics[width=0.8 \linewidth]{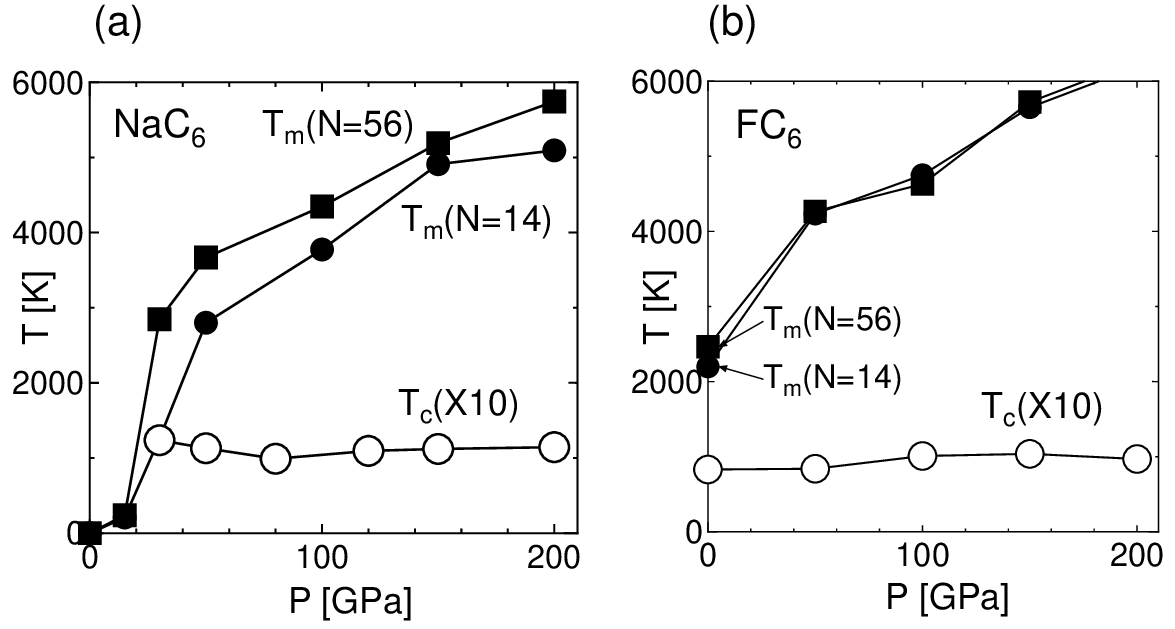}
\end{center}
\caption{
(a) $T_{\rm m}$  and  $T_{\rm c}$ of NaC$_6$ as  functions of  $P$, where   solid circles and solid squares  represent  $T_{\rm m}$ for the $N=14$   and 56 systems, respectively. Here, empty  circles  represent  $T_{\rm c}  \times 10$.
(b) $T_{\rm m}$  and  $T_{\rm c}$ of FC$_6$ as  functions of  $P$, where   solid circles and solid squares  represent  $T_{\rm m}$ for the $N=14$  and 56 systems, respectively. 
}
\label{NaC6-FC6}
\end{figure}

We found that the  $T_{\rm m}$ of NaC$_6$  is zero and unstable at $P=0$. 
However, $T_{\rm m}$ is about 2,000 K at $P=30$ GPa and increases with pressure, reaching about 5,000 K at $P = 200$ GPa.
This result seems to be consistent with those of previous works showing that   NaC$_6$  is dynamically unstable for $P \simk 30$ GPa\cite{Sano2022,Khan2022}.
On the other hand, $T_{\rm c}$ was calculated as 149, 123,  and 113 K  at $P=20$, 30, and 50 GPa, respectively.
As the pressure decreases, the melting point decreases and the system becomes unstable. 
However, $T_{\rm c}$ appears to increase with decreasing pressure for $P \simk 50$ GPa.

 Figure 6(b) shows  the $T_{\rm m}$ of FC$_6$ as a function of pressure $P$, where solid circles and  solid squares represent the results for $N=14$ and 56, respectively.
In this case, $T_{\rm m}$ is $\sim 2000$ at   $P=0$ and increases with pressure and exceeds  6000 K at  $P=200$ GPa.
Unlike NaC$_6$,   FC$_6$ is stable  even at room temperature and pressure.
$T_{\rm c}$ is 83 K at $P=0$ and increases slightly with pressure.
In a previous study, FC$_6$ was shown to be dynamically unstable by an interpolation method\cite{Sano2022}.
However,  this is not the case in the tetrahedron method and we can obtain  $T_{\rm c}$ for FC$_6$. 

\begin{figure}[ht]
\begin{center}
\includegraphics[width=0.8 \linewidth]{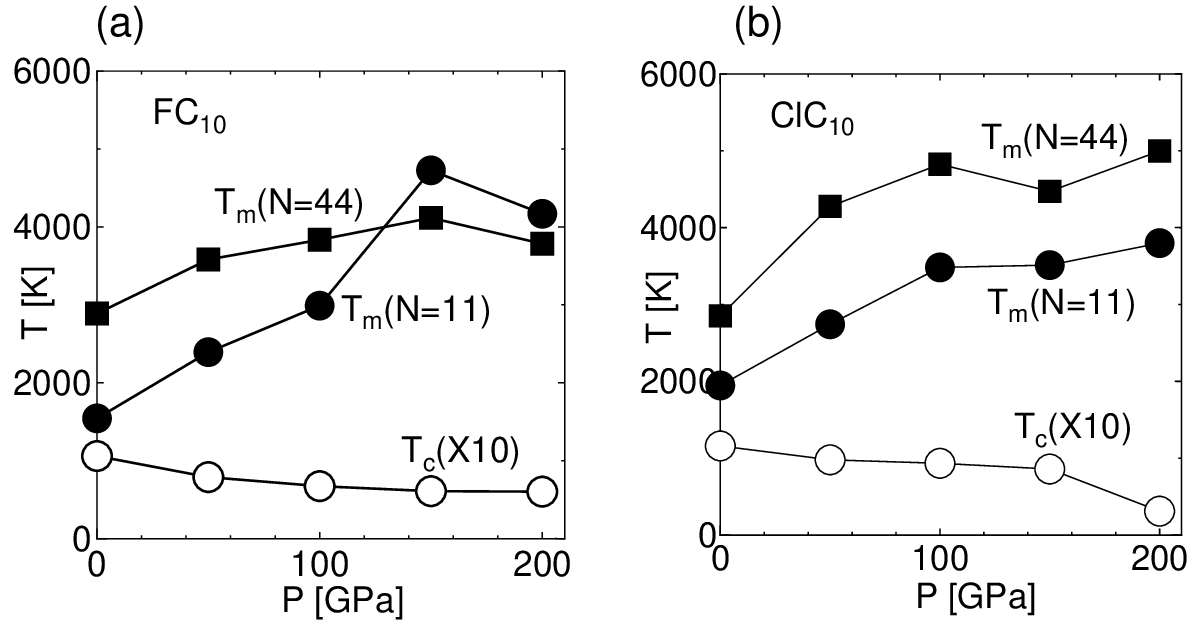}
\end{center}
\caption{(a) $T_{\rm m}$  and  $T_{\rm c}$ of FC$_{10}$ as  functions of  $P$, where  solid circles and solid squares  represent  $T_{\rm m}$ for the $N=11$  and 44 systems, respectively. Here, empty  circles  represent  $T_{\rm c} \times 10$.
(b) $T_{\rm m}$  and  $T_{\rm c}$ of ClC$_{10}$ as  functions of  $P$, where   solid circles and solid squares  represent  $T_{\rm m}$ for the $N=11$   and 44 systems, respectively.
}
\label{FC10-ClC10}
\end{figure}
Figure 7(a)  shows  the $T_{\rm m}$ of  FC$_{10}$  as a function of pressure $P$, where solid circles and  solid squares represent the results for $N=11$ and 44, respectively.
Similar to  Fig. 6, $T_{\rm c}$ as a function of pressure $P$ is also shown as empty circles.
Although the size dependence will be somewhat large in this case,  $T_{\rm m}$ may  be about 2000 K at  $P = 0$.
The figure also shows that $T_{\rm m}$ increases with pressure and reaches a maximum at about $P = 150$ GPa.
On the other hand,  $T_{\rm c}$ is about 110 K at    $P = 0$ and decreases with increasing pressure.
This material is also expected to exhibit superconductivity under normal pressure, similar to FC$_6$.

 Figure 7(b)  shows  the $T_{\rm m}$ of  ClC$_{10}$  as a function of pressure $P$, where solid circles and  solid squares represent the results for $N=11$ and 44, respectively.  Empty  circles  represent  $T_{\rm c} \times 10$.
 In this system, the behaviors of $T_{\rm m}$ and $T_{\rm c}$ are similar to those of FC$_{10}$.

For detailed data on NaC$_6$, FC$_6$, FC$_{10}$, and ClC$_{10}$, Table \ref{table1} gives  $T_{\rm m}$, $T_{\rm c}$, $\lambda$, and $\omega_{\rm log}$, where the numbers in parentheses are pressure in GPa.
Here,  $T_{\rm m}$ values are for the $N=14$ system for $X$C$_{6}$ and for the $N = 11$ system for $X$C$_{10}$.
\begin{table}[h]
\caption{$T_{\rm m}$, $T_{\rm c}$, $\lambda$,  and $\omega_{\rm log}$  for  NaC$_6$, FC$_6$,   FC$_{10}$, and  ClC$_{10}$.}
\begin{tabular}{l|cccccc} \hline
                                                 &   NaC$_6$   &   NaC$_6$     &   FC$_6$ &   FC$_6$          \\
                      $P$ [GPa]               &  (30)       &     (200)     &   (0)              &      (200)          \\     \hline    
  $T_{\rm m}$  [10$^3$K]       &    1.3        &       5.1            &    2.2            &     6.2                 \\                                                 
    $T_{\rm c}$  [K]                  &    124         &     114             &    83           &    97                \\    
 $           \lambda$                    &  3.81           &  1.76           &   1.37             &   1.45            \\ 
 $  \omega_{\rm log}$ [K]    &  407        &  744               &  731             &  830          \\     \hline   \hline
                                        &   FC$_{10}$   &   FC$_{10}$     &    ClC$_{10}$ &    ClC$_{10}$         \\        
               $P$ [GPa]               &          (0)   &   (200)         &    (0)              &        (200)          \\     \hline   
 $T_{\rm m}$  [10$^3$K]       &    1.5        &       4.2            &    1.9            &     3.8                 \\                                                              
    $T_{\rm c}$  [K]                  &    106         &     60             &    116           &    31                   \\    
 $           \lambda$                    &   1.78           &  1.13           &   2.07           &   0.67            \\ 
 $  \omega_{\rm log}$ [K]    &  697        &  653               &  662             &  997          \\     \hline   
\end{tabular}
\label{table1}
\end{table}

Results for other compounds at $P=0$ are shown in Table \ref{table2}.
For $T_{\rm m}$, we present the $N=14$ system for $X$C$_{6}$, the $N=11$ system for   $X$C$_{10}$, and  the $N=13$ system for $X$C$_{12}$.
 The table shows that $X$C$_{6}$ has a relatively high $T_{\rm c}$, but $T_{\rm m}$ is not  high.
On the other hand, $X$C$_{10}$ and $X$C$_{12}$  have high $T_{\rm m}$, but $T_{\rm c}$ is low.
These results suggest that $T_{\rm c}$ decreases as the proportion of $X$ atoms in the system decreases.
This can be interpreted as a reduction in the number of effective carriers  that the $X$ atom brings to the system\cite{Sano2022}.
This leads to a decrease in $T_{\rm c}$.
Conversely, an increase in the number of  effective carriers reduces the stability of the system and decreases $T_{\rm m}$.
\begin{table}[h]
\caption{$T_{\rm m}$, $T_{\rm c}$, $\lambda$,  and $\omega_{\rm log}$  for other compounds at $P=0$.}
\begin{tabular}{l|cccccc} \hline
                                     &   HC$_6$   &   LiC$_6$     &   BC$_6$      &   OC$_6$        &   ClC$_6$   &   HC$_{10}$     \\     \hline                                                 
 $T_{\rm m}$ [10$^3$K]      &  0.30        &  0.11               &  0.17             &  0.51            &  0.18       &     2.2          \\ 
    $T_{\rm c}$  [K]                  &    43         &     57                 &    20           &    52              &   73          &    14           \\        
$           \lambda$                    &   0.75          &  3.47         &   2.37           &   1.23           &     1.14         &  0.53     \\ 
 $  \omega_{\rm log}$ [K]    &  889        &  209               &  113             &     532        &      825         &     902  \\     \hline   
\hline  
                                                &  OC$_{10}$  &  NaC$_{10}$   &    BC$_{12}$  &    NaC$_{12}$   &   FC$_{12}$  &    ClC$_{12}$     \\     \hline                                                 
 $T_{\rm m}$ [10$^3$K]      &    1.5            &  1.3                   &    0.57            &    1.2                &       3.0            &    1.7             \\ 
    $T_{\rm c}$  [K]              &      32            &    13                      &     6                &   32                &     17                &      21             \\    
$           \lambda$                    &   0.95          &  0.57           &   0.53                   &   0.73             &   0.58             &    0.60   \\ 
 $  \omega_{\rm log}$ [K]    &  463        &  670               &  336                     &  765               &       955        &   951      \\     \hline   
\end{tabular}
\label{table2}
\end{table}
%
\begin{figure}[hb]
\begin{center}
\includegraphics[width=0.8 \linewidth]{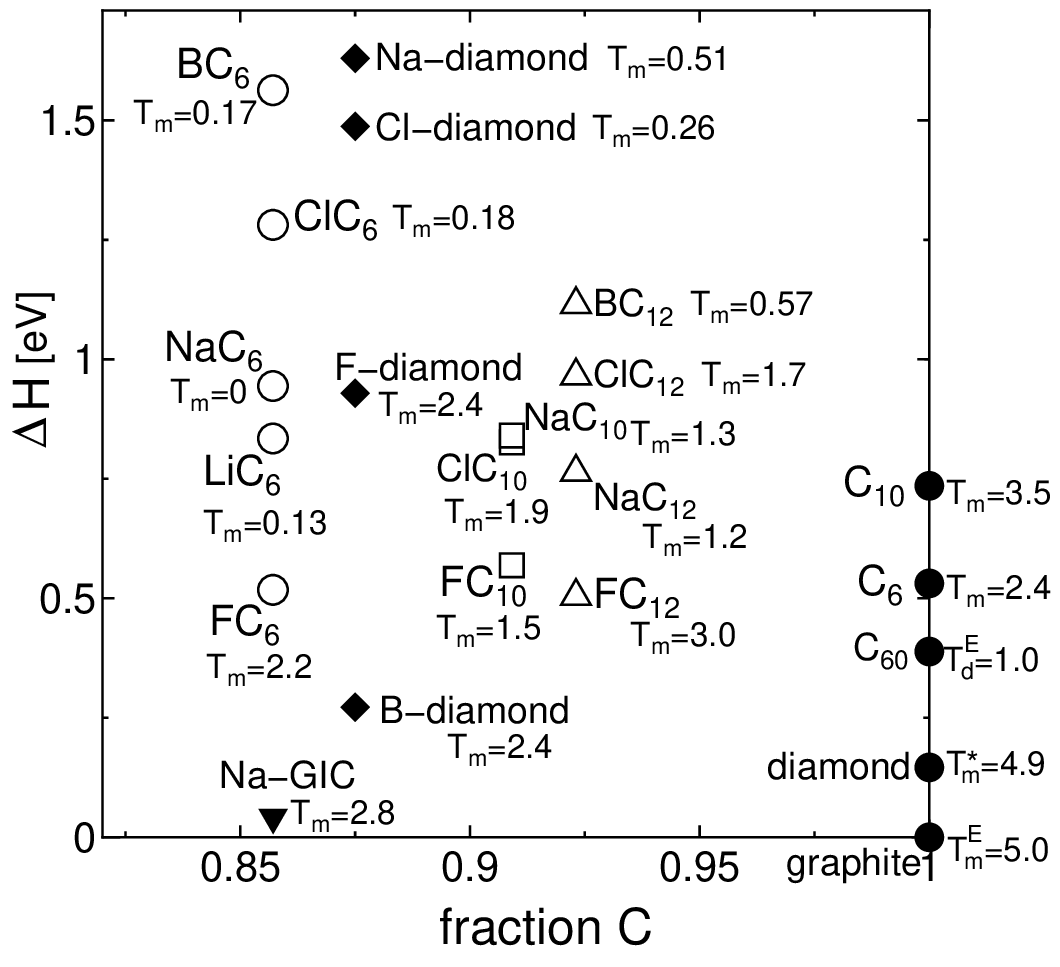}
\end{center}
\caption{ Formation enthalpies  $\Delta H$ as a function of the fraction of carbon in  compounds with $T_{\rm m}$, where the unit of $T_{\rm m}$ is $10^3$ K.
Here,   $X$-diamond means that one atom is replaced by an $X$-atom in  the 8-site diamond cell, and Na-GIC represents
 a graphite intercalation compound with sodium between the layers.
}
\label{P-lam}
\end{figure}
\subsection{ Formation enthalpies  at $P=0$ and $T=0$}\label{XC6}
From the perspective of actually synthesizing these compounds, not only  $T_{\rm m}$ but also the formation enthalpy  may be important.
We defined the  formation enthalpy  $\Delta H$   such as $\Delta H = H_{\rm NaC_6}- (H_{\rm Na}+6H_{\rm gra})/7$ for NaC$_6$, 
and similarly for other compounds.
Here,  $H_{\rm NaC_6}$,  $H_{\rm Na}$, and $H_{\rm gra}$ are the  enthalpies  at  $T=0$  per atom  of  NaC$_6$,  sodium  in a bcc structure, and  graphite, respectively.
Note that the definition of $\Delta H$ given here is insufficient to estimate an exact value of the formation  enthalpy.
To obtain this,  it is necessary to compare the energies of all compounds that contain the elements in the target compound in any ratio.
Since this is very difficult  in practice,  we estimated $\Delta H$ as information that is useful for synthesizing compounds.

 Figure 8  shows the  $\Delta H$ of $X$C$_6$,   $X$C$_{10}$, and  $X$C$_{12}$ at $P=0$, where the horizontal axis is the fraction of carbon contained in  compounds. Here, $\Delta H$ of $X$C$_6$ is represented by empty circles,  that of $X$C$_{10}$  by empty squares, and   that of $X$C$_{12}$  by empty triangles.
In addition, there is  the $\Delta H$ of $X$-atom-doped diamond  (denoted as $X$-diamond), in which one carbon atom in a diamond supercell made up of eight atoms is replaced by an $X$-atom. These are represented by solid diamonds in the figure.
The solid triangle stands for the result of the graphite intercalation compound that has sodium between the layers (Na-GIC) \cite{Hao2023a}.
Each point of $\Delta H$ is accompanied by a value of $T_{\rm m}$ in units of $10^3$ K.
The  $T_{\rm m}$ values for $X$C$_6$, $X$C$_{10}$, and $X$C$_{12}$ correspond to those for the systems with N = 14, 11, and 13, respectively.

For reference,   the  formation enthalpies per atom  of graphite,  pure diamond, C$_6$, C$_{10}$, and   C$_{60}$ have also been added, where these are represented by solid circles.
Here, $T^{\rm E}_{\rm m}$ stands for the experimental melting temperature, $T^{*}_{\rm m}$ is the melting temperature at the triple point near 12 GPa, and $T^{E}_{\rm d}$ is the decomposition temperature observed in the experiment\cite{Sundar1992}.
The  results for $X$C$_6$ and  $X$C$_{12}$   almost agree with those of the others\cite{Wei-2016}.

Comparing the difference in the value of $\Delta H$ with those in other studies\cite{Material-project}, we found that it is about 0.008 eV for diamond and about 0.012 eV for C$_{60}$.
For compounds NaCl, B$_4$C, and Be$_2$C\cite{Material-project}, this difference is about 0.16, 0.005, and 0.11 eV, respectively, although not shown in the figure.
For $X$C$_6$ and $X$C$_{12}$,  the discrepancy between our results and others' is roughly 0.06 eV \cite{Wei-2016}.
In the case of  compounds, the discrepancy of $\Delta H$ does not seem to be very small.
%

The figure shows that the $\Delta H$ of  C$_{60}$   is  about 0.4 eV and  close to that of  C$_6$.
C$_{60}$ is a  material well known as fullerene and can  now be  synthesized.
Therefore, it is expected that the synthesis of C$_6$ is possible.

Unfortunately,  $X$C$_6$ has low $T_{\rm m}$,  except for  FC$_{6}$ and OC$_{6}$,  and may  be unstable at room temperature.
On the other hand, $X$C$_{10}$ and  $X$C$_{12}$ have higher $T_{\rm m}$ than $X$C$_{6}$,
even though their $\Delta H$ values are similar to each other.
The value of $\Delta H$ for B-diamond is  about 0.25 eV, which is small, where its $T_{\rm c}$ is calculated as 32 K.
This result seems to be related to the fact that boron-doped  diamond was found to be a superconductor in  experiments.\cite{Kawano,Okazaki,Ekimov2004}
 On the other hand,  $\Delta H$ values of F-, Cl-, and Na-diamonds are around 1 eV or larger. 
This might correspond to the fact that   superconducting diamonds doped with these elements have not been found.

The Na-GIC (sodium intercalated graphite)\cite{Hao2023a} considered here has the same crystal structure as the experimentally discovered Li-GIC (lithium intercalated graphite)\cite{Material-project}.
This type of Na-GIC has not been found experimentally.  
However, it is an interesting isomer of NaC$_6$ with the sodalite structure, since    finite $T_{\rm c}$ is found  theoretically.\cite{Hao2023a}
%
\begin{figure}[thb]
\begin{center}
\includegraphics[width=0.8 \linewidth]{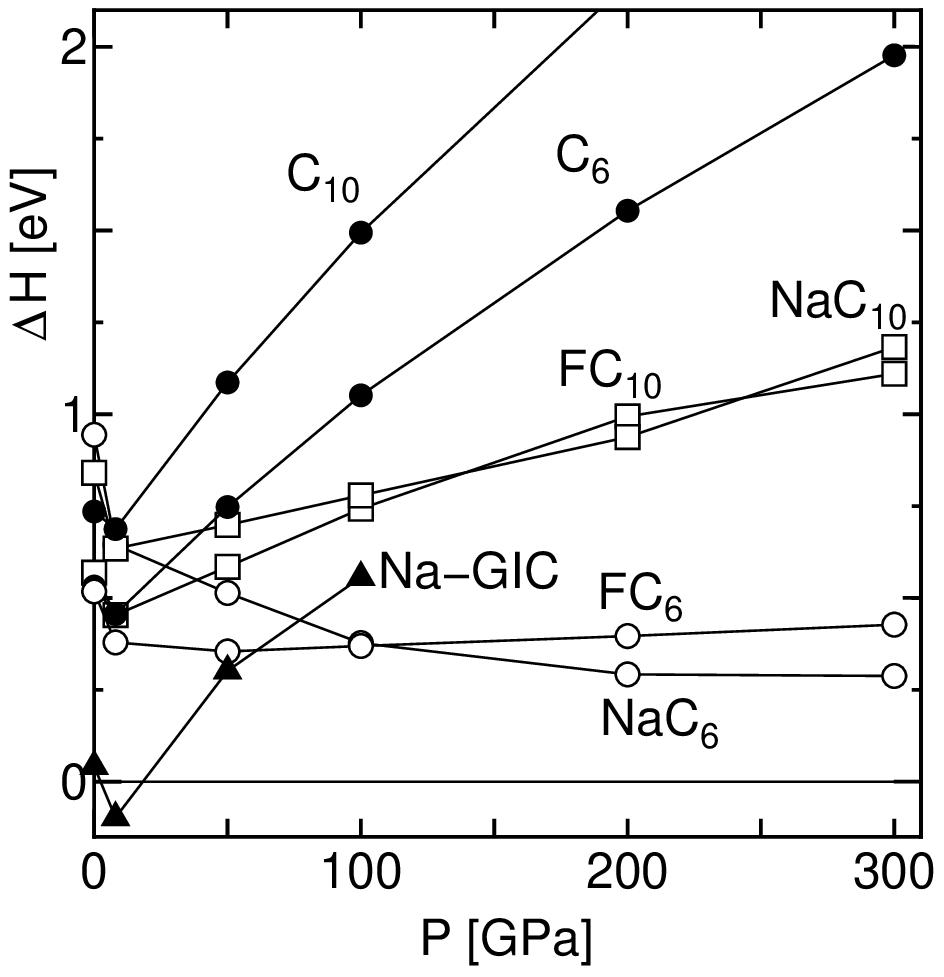}
\end{center}
\caption{
  $\Delta H$ as a function of  $P$ for C$_6$, C$_{10}$,  FC$_6$, NaC$_6$, FC$_{10}$,  NaC$_{10}$, and  Na-GIC.
  }
\label{P-B}
\end{figure}
\subsection{ Formation enthalpies  as a function of $P$ at $T=0$}
The pressure dependence of formation enthalpies is also interesting.
Figure 9   shows   $\Delta H$ as a function of $P$ for C$_6$, C$_{10}$, FC$_6$, NaC$_6$, FC$_{10}$, NaC$_{10}$, and  Na-GIC.
Here, the $\Delta H$ values are calculated relative to diamond, except when $P=0$.
This is because the enthalpy of diamond is lower than that of graphite at pressures above about 2 GPa.
Furthermore, fluorine and sodium undergo structural phase transitions when the pressure is increased\cite{Christian2022,Christensen2001}, so the $\Delta H$  of the compounds must be calculated using the enthalpies corresponding to their structures.

For fluorine, the $C2/c$ structure is used for $P<20$ GPa, and the $Cmca$ structure is used otherwise.\cite{Christian2022}
For sodium,  the bcc, fcc,  $Cs$IV, and  $P6_3/mmc$  structures are used for $P<60$, $P=100$, 200, and 300 GPa,  respectively.\cite{Christensen2001}

Figure 9  indicates that    $\Delta H$ increases with $P$  for C$_6$,  C$_{10}$,    FC$_{10}$, NaC$_{10}$, and  Na-GIC.
These substances have  relatively low density, so their enthalpy increases under high pressure.
In fact,  the densities of  diamond, C$_6$, and  C$_{10}$ are 4.64, 3.97, and 3.79 ${\rm g/cm^3}$, respectively, at $P=200$ GPa.
Furthermore, those of NaC$_6$ and Na-GIC are 4.18 and 3.54 ${\rm g/cm^3}$, respectively,  at $P=100$ GPa.
Here, the melting point of Na-GIC is 0 K at $P \sim 110$ GPa, and it may be thermally unstable at pressures above this.

The figure also indicates that the  $\Delta H$  of NaC$_6$ decreases with increasing $P$, while the others  generally increase with $P=0$ except FC$_6$. The $\Delta H$  of FC$_6$ seems to be almost independent of $P$. 
We have confirmed that the  $\Delta H$ values  for NaC$_6$ and FC$_6$ are almost the same  at $P=300$  and 400 GPa.
Therefore, it seems that these  $\Delta H$ values will not decrease even if the pressure is increased further.

In contrast to NaC$_6$, the $\Delta H$ of NaC$_{10}$ increases with pressure. 
This can be interpreted as the crystal structure of NaC$_{10}$ being more sparse than that of NaC$_6$.
This  may  also apply to  the relationship between FC$_6$  and FC$_{10}$.
%
\subsection{Formation free energies for  NaC$_6$ and  FC$_6$}\label{XC6}
As shown in  Fig. 9,  the $\Delta H$ of NaC$_6$ decreases with increasing pressure and becomes about 0.29 eV at $P = 300$GPa.
In this case,  the effects of  finite temperature, such as lattice vibrations, may not be negligible near the melting point.
To  clarify  the effects of phonons, we  calculated the Helmholtz free energy $F(T)$ by  harmonic approximation.
Comparing $F(T)$ with experimental results\cite{Huang2022}, we introduced a phenomenologically corrected  free energy $\tilde{F}(T)$. 
It reproduces the experimental results better than $F(T)$ and allows us to estimate the free energy of the liquid state.
By adding $\tilde{F}(T)$ to the enthalpy at $T=0$, we can approximately obtain the Gibbs free energy at finite temperatures.
Detailed calculation methods are given in Appendix B.
%
\begin{figure}[thb]
\begin{center}
\includegraphics[width=0.8 \linewidth]{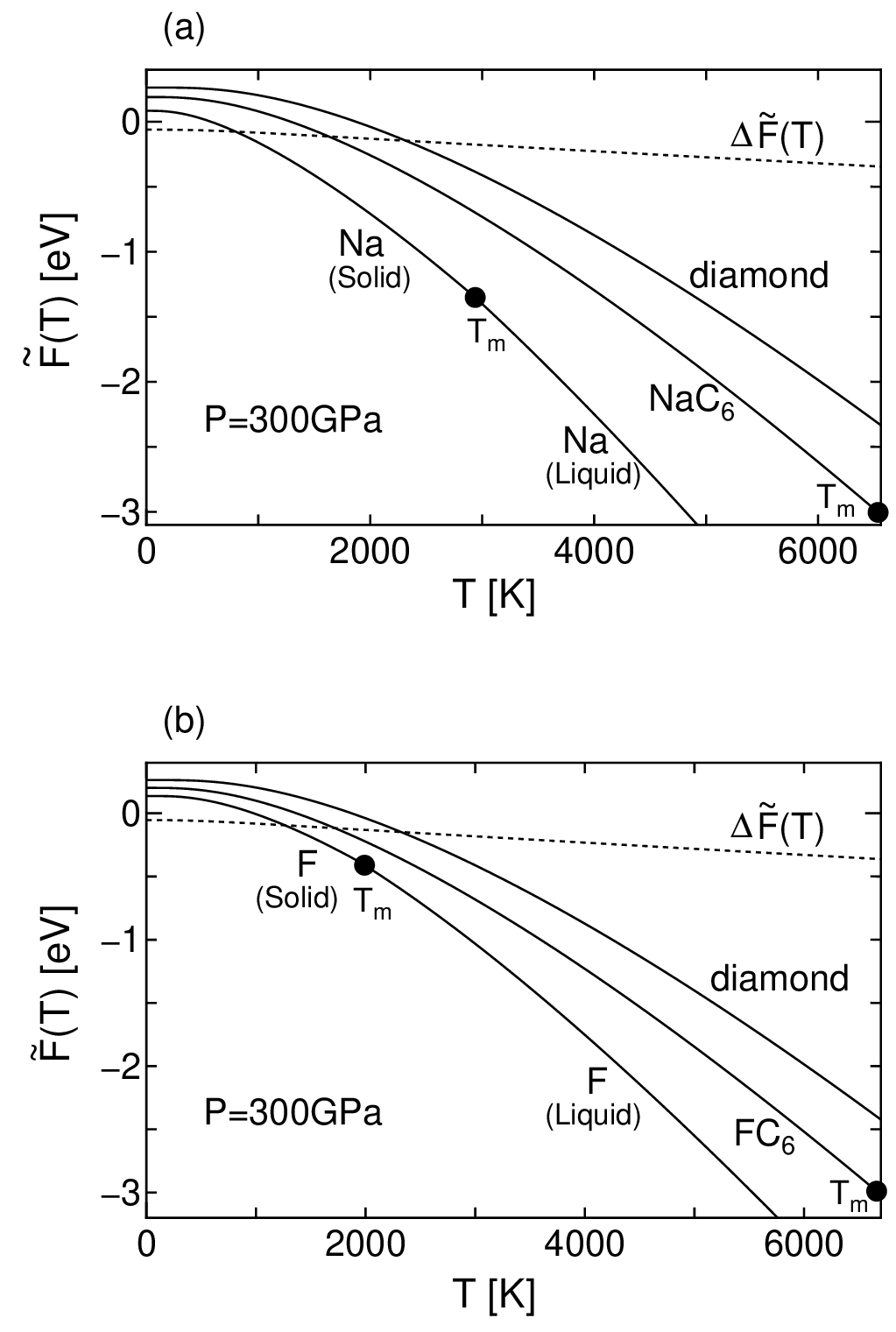}
\end{center}
\caption{Free energy as a function of temperature for  (a) Na,   NaC$_6$,   and diamond, and  (b)  F,   FC$_6$,   and diamond at $P=300$ GPa. 
Here, solid circles represent $T_{\rm m}$.
The dotted lines show  the formation free energy $\Delta \tilde{F}(T)$, which was determined using a similar definition as for  $\Delta H$.
 }
\label{fig10}
\end{figure}
%

Figure 10(a) shows $\tilde{F}(T)$ as a function of $T$ for diamond, NaC$_6$, and sodium at $P=300$ GPa.
We also present the  formation   free energy, $\Delta \tilde{F}(T)$, which is defined as $\Delta  \tilde{F}(T) = \tilde{F}_{\rm NaC_6}(T)- (\tilde{F}_{\rm Na}(T)+6\tilde{F}_{\rm dia}(T))/7$. 
Here, $\tilde{F}_{\rm NaC_6}(T)$, $\tilde{F}_{\rm Na}(T)$, and  $\tilde{F}_{\rm dia}(T)$ are the  $\tilde{F}(T)$  per atom for   NaC$_6$, sodium, and diamond, respectively.
These values decrease with increasing temperature owing to entropy.
As shown in the figure, $\Delta \tilde{F}(T)$ is about $-0.34$ eV at $T=6550$ K, where this temperature corresponds to the $T_{\rm m}$ of NaC$_6$, and the $T_{\rm m}$  of diamond is about 9300 K under the same pressure.
Here, to obtain  $T_{\rm m}$,  we used the $N=$14 system for NaC$_{6}$ and the $N=8$ system for diamond.

Recalling that $\Delta H \simeq 0.29$ eV,  the value of $\Delta H + \Delta \tilde{F}(T)$ becomes about $-0.05$ eV.
Since the value is negative,  the formation of NaC$_ 6$ under this condition can be expected.\cite{NaC6}
The result also suggests that high pressure and high temperature may favor the synthesis of  NaC$_6$.

By applying  a similar analysis as described above to FC$_6$,  $\Delta \tilde{F}(T)$   is found to be about $-0.36$ eV at $T=6660$ K, as shown in Fig. 10(b). Here, this temperature corresponds to   $T_{\rm m}$  at $P=300$ GPa.
We assumed that $c_{\rm solid}$=1.06 for fluorine, FC$_6$, and diamond. For the liquid state of  fluorine,  $c_{\rm liquid}$ was assumed to be 1.2.
Since the $\Delta H$ of  FC$_6$   is about 0.43 eV  at $P=300$ GPa, we obtained $\Delta H + \Delta \tilde{F}(T) \simeq 0.07$ eV. 
This value is  small but positive. Thus, the synthesis  of FC$_6$ may be more difficult than that of NaC$_ 6$.

In the above discussion,  we have not addressed the  contribution of electrons at finite temperatures.
To consider it, we estimated the free energy of electrons.
At $P=300$ GPa, NaC$_6$ is metallic and the value of DOS at the Fermi level, $D_{\rm F}$, is 0.204 states/eV/atom.
Using the free electron model, we calculated the contribution of free energy to be  $\sim -\frac{\pi^2}{6}(k_BT)^2D_{\rm F}$  at low temperatures. The value is  about $-0.11$  eV at $T=6550$ K.
If liquid sodium is in the metallic state, the value of the free energy  is expected to be of the same order as  NaC$_6$.
On the other hand, we found that diamond has an energy gap of 5.6 eV at $P=300$ GPa, and  the electron contribution to the free energy is negligible.
In this case, the $\Delta \tilde{F}(T)$ of NaC$_ 6$ may decrease by about 0.09 eV.

However, we used $\tilde{F}(T)$, which closely reproduces the experimental results. 
Since the experimental results should include the effect of  electrons at finite temperatures, $\tilde{F}(T)$ should be considered to include the effect of electrons, even if it is not complete.
If so, adding the electron contribution to  $\tilde{F}(T)$  could result in double counting.
Similarly, the effects of volume changes at finite temperatures may already be included.
In any case, the contribution of these effects may be small and   would not significantly change the results.
%
\section{Summary}
 By combining the FPMD method and the Z-method, we investigated the $T_{\rm m}$ of carbon compounds $X$C$ _6$, $X$C$ _{10}$, and $X$C$ _{12}$ with the sodalite structure.
 To verify the validity of the method, we applied it to the diamond system and compared the results with those of other theoretical  and experimental works.
We also discussed the assumptions adopted in the Z-method by comparing them with experimental results.
In addition to  the melting points, we  obtained the superconducting transition temperature  $T_{\rm c}$ by the optimized tetrahedron method in QE.

These results  suggest that  FC$ _6$,  FC$ _{10}$, and ClC$ _{10}$  can exist stably at sufficiently  high temperatures ($T \simj 1000$ K)  even under normal pressure, and the $T_{\rm c}$ values of these compounds reach nearly 100 K.
Although the $T_{\rm m}$ of NaC$ _6$ drops rapidly at pressures below 30 GPa, $T_{\rm c}$ is over  120 K for $P \simj 30$ GPa.
We also showed results for  several other compounds that have relatively low $T_{\rm c}$  but are stable at  ambient  pressure.
If these compounds can actually be synthesized, it would be possible to observe high-temperature superconductivity mediated by phonons.

To examine the feasibility of these compounds, we  calculated the  formation  enthalpies $\Delta H$ from the ground state energies of the elements. Unfortunately, all of these values for compounds were positive and above 0.5 eV at $P=0$.
This suggests that the synthesis of these compounds is difficult.
We also examined the pressure dependence of $\Delta H$ for  C$ _6$, C$ _{10}$, and several  representative compounds.
Since C$ _6$ and C$ _{10}$ have lower densities than diamond, $\Delta H$ increases with pressure.
However, the $\Delta H$ of NaC$_{6}$ decreases with increasing pressure, from about 0.94 eV at $P=0$ to 0.29 eV at $P=300$ GPa.

In this case, the effect of finite temperature may make the Gibbs formation free energy negative.
To clarify this, we introduced a phenomenologically corrected formation free energy $\Delta \tilde{F}(T)$. 
Applying this to NaC$_{6}$, we found  the value of $\Delta \tilde{F}(T)$ to be approximately $-0.36$ eV at $T=6600$ K and $P=300$ GPa.
This cancels out the positive $\Delta H$, and $\Delta \tilde{F}(T) + \Delta H$, which corresponds to the Gibbs formation free energy, becomes negative.

This result suggests that sufficiently high temperatures and pressures may be effective for the formation of NaC$_6$.
Applying a similar analysis to FC$_6$,  we found that the Gibbs formation free energy is small but positive under conditions similar to those for NaC$_6$.
This result indicates that the synthesis of FC$_6$ may be more difficult than that of NaC$_6$.

\section*{ACKNOWLEDGMENTS}
This work was  supported by JSPS KAKENHI Grant Number  JP19K03716.

%
%

 \appendix{ }  \section{Simulation Details}

All calculations of FPMD and electronic structures were carried out by using the density functional theory of Perdew-Burke-Ernzerhof (PBE) generalized gradient approximation and the projector augmented wave pseudopotential as implemented in QE package.\cite{QE} 
The cutoff energies for the wave functions and charge densities were set to the suggested minimum  values given in the pseudopotential files.\cite{QE}
For example, for a carbon atom, the two  values are 40 and 323 Ry, respectively.
Convergence thresholds were set as  $10^{-4}$ a.u. in  total energy and  $ 10^{-3}$ a.u. in force.

The $\bm {k}$  of $6 \times 6 \times 6 $ mesh  in  the first Brillouin zone was used  for  diamond systems  with $N=$8, $X$C$_6$ systems with $N=$14, and $X$C$_{10}$ systems with $N=$11.
For other systems with $N$ less than 30, a   $4 \times 4 \times 4 $ mesh was used.
For all systems with  $N$ more than 30, a $2 \times 2 \times 2 $  mesh was used.
 In the calculation of  FPMD,  we used the  `vc-md` option \cite{QE} in QE, which uses the Beeman algorithm to integrate Newton's equation. 
The  time step is chosen to be one femtosecond (fs).

A new simulation is started with the equilibrium atomic positions at $T=0$, where  these atomic positions   are obtained by structural optimization under the target pressure. 
The initial velocity distribution is a Boltzmann distribution determined by the target temperature.
The instantaneous temperature $T_{\rm i}$  is calculated at the end of every atomic move,  and the temperature of the system  is rescaled to the target temperature at each step.
The instantaneous pressure $P_{\rm i}$ of the system changes with each step, but is adjusted to the target pressure by modifying the cell volume.
The root mean square  of the displacement of the atomic positions from their initial  positions at $t=0$  corresponds to the Lindemann parameter $L_{\rm p}$.
Here, $L_{\rm p}$, $P_{\rm i}$, and $T_{\rm i}$ are recorded every 10 steps  during the simulation.

\begin{figure}[thb]
\begin{center}
\includegraphics[width=0.8 \linewidth]{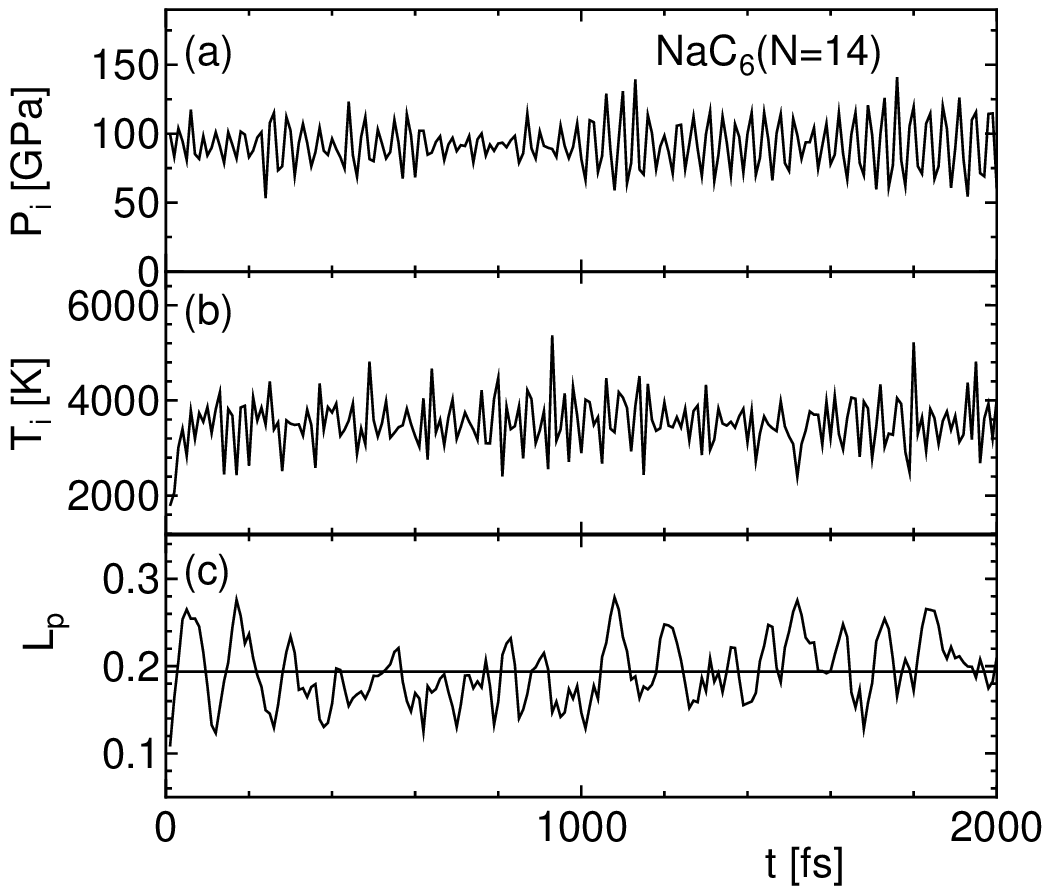}
\end{center}
\caption{(a) Instantaneous pressure $P_{\rm i}$ as a function of $t$, (b) instantaneous temperature $T_{\rm i}$   as a function of $t$, and (c)  Lindemann parameter $L_{\rm p}$ as a function of $t$, where a straight line  represents  the average value of  $L_{\rm p}$.
The system  used  is NaC$_6$ ($N=$ 14), and the  target pressure and   target temperature are set to 100 GPa and 3500 K, respectively. 
 }
\label{fig-a1}
\end{figure}
%

\begin{figure}[tbh]
\begin{center}
\includegraphics[width=0.8 \linewidth]{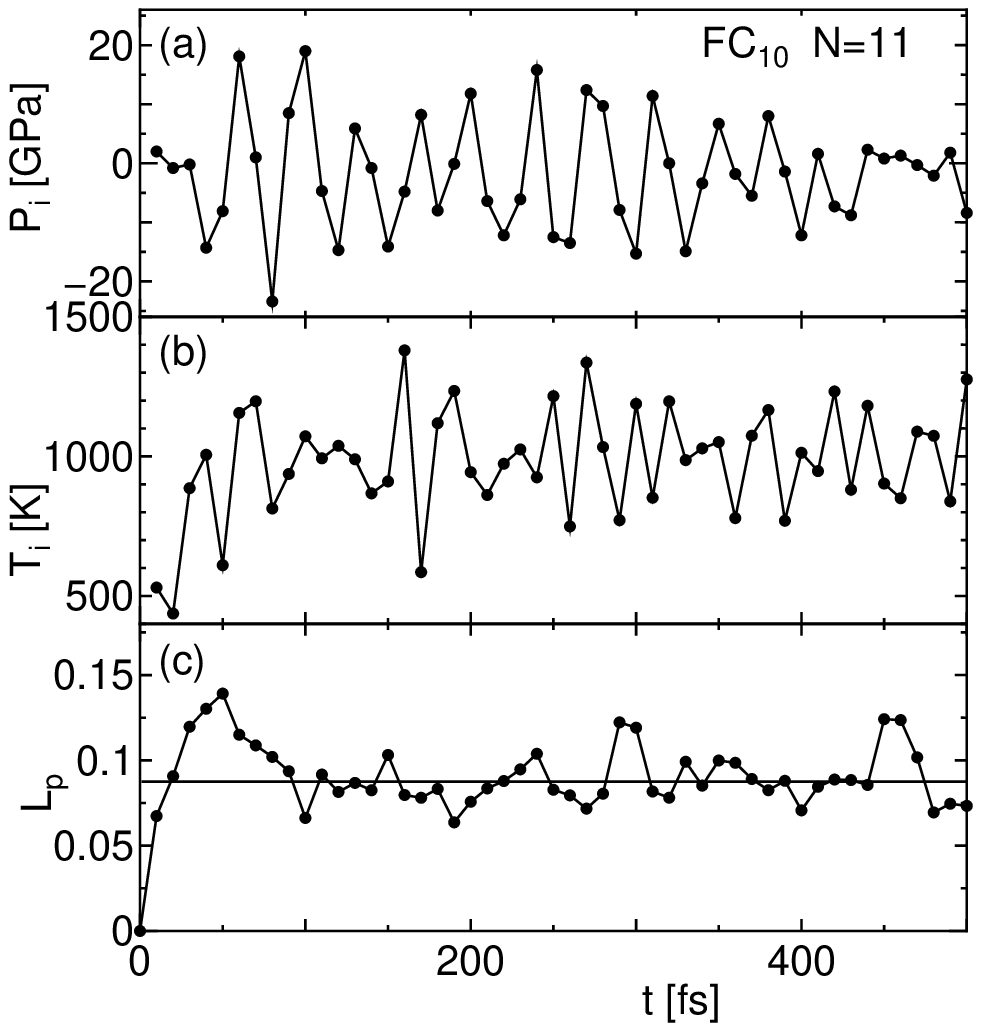}
\end{center}
\caption{
(a)  Instantaneous pressure $P_{\rm i}$ as a function of $t$, (b)  instantaneous temperature $T_{\rm i}$   as a function of $t$, and (c) the Lindemann parameter $L_{\rm p}$ as a function of $t$, where a straight line represents  the average value of  $L_{\rm p}$.
Here, the system is FC$_{10}$ with $N=$ 11. The target pressure and   target temperature are set to 0 GPa and 1000 K, respectively. 
}
\label{fig-a2}
\end{figure}

 Figure \ref{fig-a1}  shows the simulation results for (a) $P_{\rm i}$(a),  (b) $T_{\rm i}$, and  (c) $L_{\rm p}$ for NaC$_6$ ($N= 14$) as a function of $t$ up to 2000 fs, where the target pressure and    target temperature are set to 100 GPa and 3500 K, respectively.
 In Fig. \ref{fig-a1}(c), the straight line represents the average value of  $L_{\rm p}$.
Owing to the small system size, $P_{\rm i}$ and $T_{\rm i}$ show large fluctuations.
However, these values are around the target pressure and the target  temperature at all times, and no systematic deviations are observed within this calculation.
 The value of $L_{\rm p}$ also appears to fluctuate around the average value.

As shown in Fig. \ref{NaC6-FC6} in the main text, $T_{\rm m}$ and $T_{\rm u}$ were estimated to be approximately 3800  and 4600 K, respectively.  
The peaks of $T_{\rm i}$ in  Fig. \ref{fig-a1}(b) appear to reach $\sim 5000$ K and  exceed the above temperatures.
However, the system  maintains its crystal structure.
It is interesting that very large  instantaneous fluctuations do not necessarily lead to melting.

Figure \ref{fig-a2} shows  the simulation results for (a) $P_{\rm i}$, (b) $T_{\rm i}$, and (c) $L_{\rm p}$ for  FC$_{10}$ ($N=$11) 
as a function of $t$ up to 500 fs, where the target pressure and   target temperature are set to 0 GPa and 1000 K, respectively.
In Fig. \ref{fig-a2}(c),  the straight line represents the average  value up to 2000 fs.
We  confirmed that these values  fluctuate around the average value up to  2000 fs.

Similar to the case of NaC$_6$,  the fluctuations are large.  
As shown in the figures, it is difficult to distinguish between fluctuations in the equilibrium state and relaxation processes from the initial state  at $t=0$.
Here, we considered the relaxation process by focusing on the behavior of  $L_{\rm p}$   in Fig. \ref{fig-a2}(c).
In our simulation, the value of $L_{\rm p}$   starts  from zero at $t=0$.
The value  appears to approach the equilibrium value after $\simeq 100$ fs.
Judging from this behavior, the relaxation time may be considered to be about 100 fs. 

\begin{figure}[thb]
\begin{center}
\includegraphics[width=0.8 \linewidth]{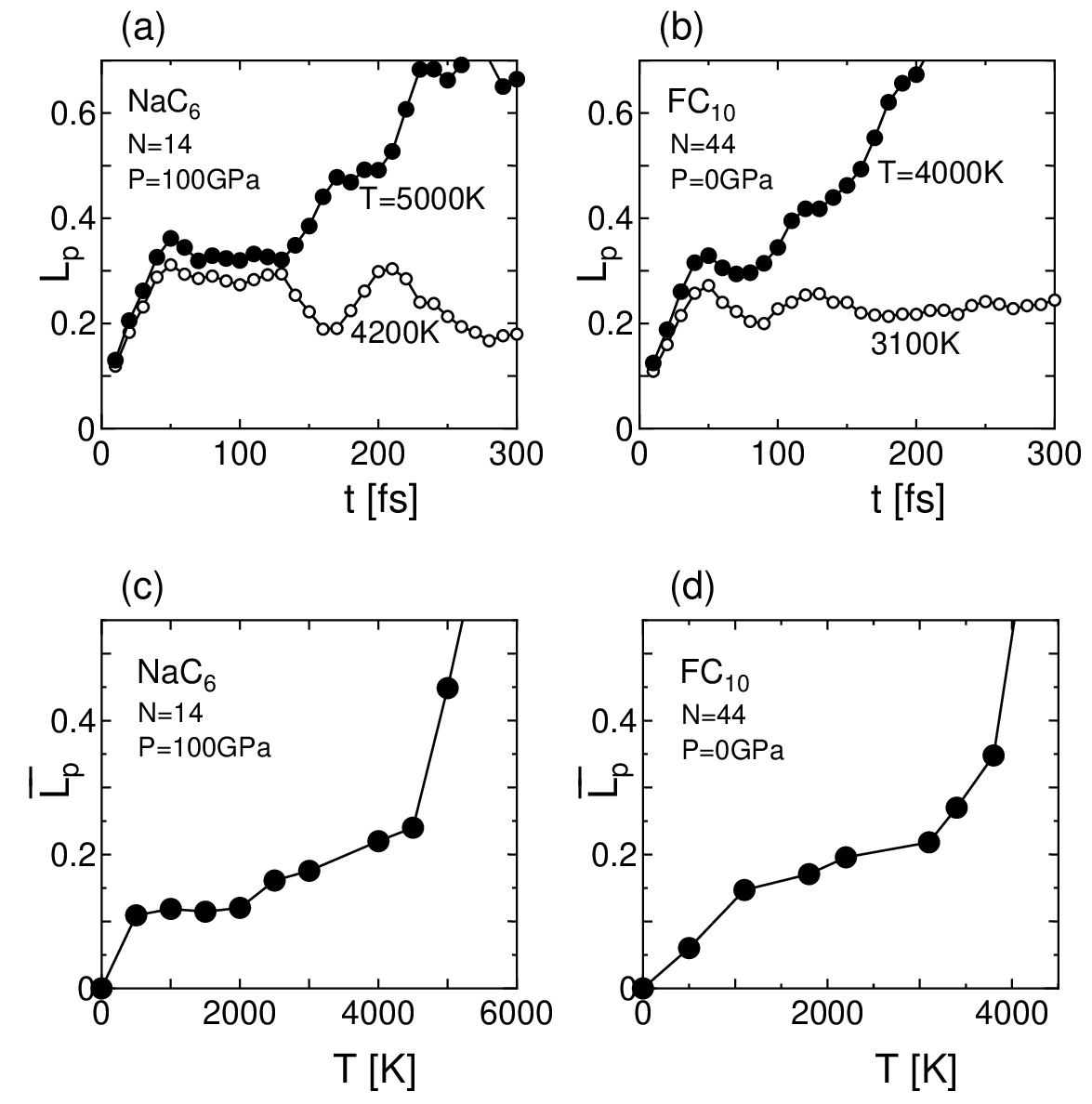}
\end{center}
\caption{ (a) $L_{\rm p}$  as a function of  $t$  for  NaC$_{6}$ at   the target temperature  $T=5000$ K  (solid circles)  and $T=4200$ K  (empty circles) with $N=14$ at   $P=100$ GPa.   (b)  $L_{\rm p}$  as a function of  $t$  at $T=4000$ K (solid  circles) and   $T=3100$ K (empty circles)  for   FC$_{10}$ with $N=44$  at  $P=0$ GPa.  
(c) $\bar{L_p}$ as a function of $T$  for    NaC$_6$ with $N=14$ at $P=100$ GPa.   (d)  $\bar{L_{\rm p}}$ as a function of $T$  for   FC$_{10}$ with $N=44$  at  $P=0$ GPa. 
 }
\label{fig-a3}
\end{figure}
If the crystal structure of a system becomes unstable at high temperatures, the value of $L_{\rm p}$ is expected to become very large.
Figures \ref{fig-a3}(a) and \ref{fig-a3}(b) show $L_{\rm p}$ as a function of $t$ for (a) NaC$_{6}$ ($N=14$, target pressure $P$=100 GPa) and (b) FC$_{10}$ ($N=44$, target pressure $P=0$ GPa).
For the case of NaC$_{6}$, $L_{\rm p}$ increases sharply with $t$ for $t \simj 150$ K at $T=5000$ K, but it seems to be almost constant at $T=4200$ K.
For the case of FC$_{10}$,  similar to  NaC$_{6}$, $L_{\rm p}$ is nearly constant at $T=3100$ K, but seems to increase rapidly with $t$  at   $T=4000$ K.
These results suggest that the system becomes unstable  at  $T \sim 4600$ K for NaC$_{6}$, and  at  $T \sim 3500$ K for   FC$_{10}$.

To calculate $T_{\rm u}$ more systematically, we used $\bar{L_{\rm p}}$, which is defined as the average value of $L_{\rm p}$ at $t=180$, 190, 200, and 210 fs.
Figures \ref{fig-a3}(c) and  \ref{fig-a3}(d) show $\bar{L_{\rm p}}$ as a function of the target temperature $T$ for (c) NaC$_{6}$  and (d) FC$_{10}$.
The point where the temperature dependence of $\bar{L_{\rm p}}$ changes suddenly may correspond to $T_{\rm u}$.
This behavior of $\bar{L_{\rm p}}$ is the same as that observed in the diamond system discussed in the main text.
For convenience, we considered that the temperature at which  $\bar{L_{\rm p}}= 0.3$ approximately gives  $T_{\rm u}$.
In this simulation, the target temperature is varied in intervals between approximately 1/5 and 1/10 of $T_{\rm u}$, and  the temperature at which $\bar{L_{\rm p}} = 0.3$ was found by interpolation.

Although  the step width of the target temperature is  coarse  and the system used  is not so large, it may be successful in giving an approximate value of $T_{\rm u}$.
In fact,  the values of $T_{\rm m}$ estimated from our $T_{\rm u}$ roughly reproduced the experimental results, as shown in Fig. \ref{dia-souzu} in the main text and  Table \ref{tableA1} in  Appendix B.

\begin{figure}[thb]
\begin{center}
\includegraphics[width=0.8 \linewidth]{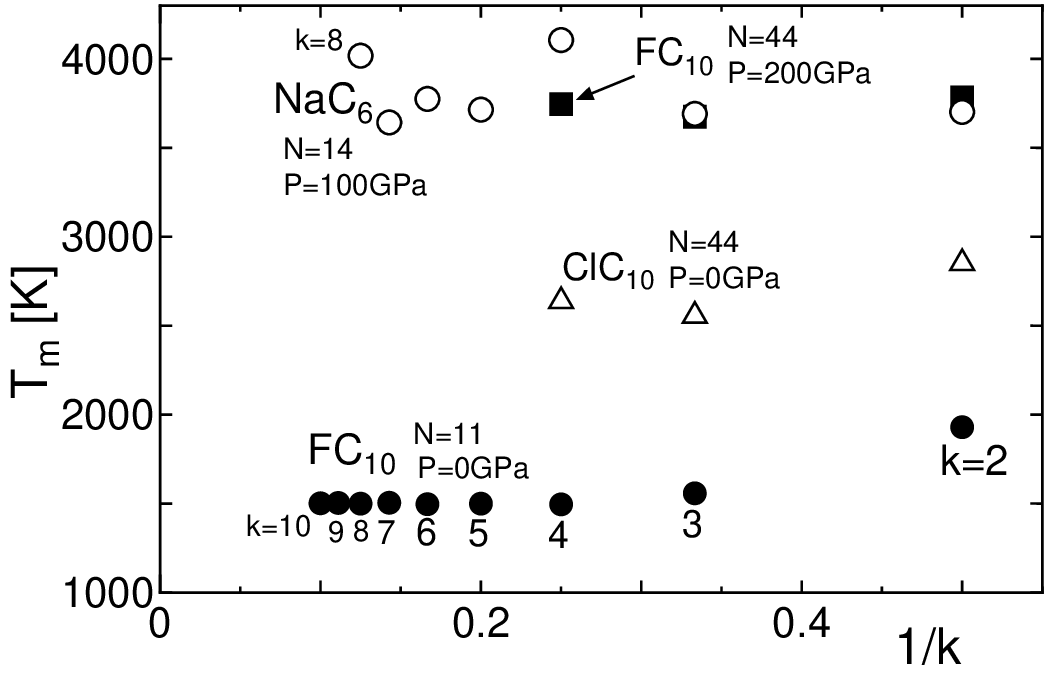}
\end{center}
\caption{  $k$-dependence of $T_{\rm m}$ for NaC$_6$ for $N=14$ at $P=100$ GPa (empty circles),  
 FC$_10$ for $N=44$  at $P=200$ GPa (solid squares),  ClC$_10$ for $N=44$  at $P=0$ GPa (empty triangles),  
and FC$_10$ for $N=11$ at $P=0$ GPa  (solid circles).
 }
\label{fig-a4}
\end{figure}
 Figure \ref{fig-a4}  shows the $k$-dependence of  $T_{\rm m}$ for  the systems of NaC$_6$ for $N=14$ at $P$=100 GPa (empty circles),   FC$_{10}$ for $N=44$  at $P=200$ GPa (solid squares),  ClC$_{10}$ for $N=44$  at $P=0$ GPa (empty triangles),  and FC$_{10}$ for $N=11$ at $P=0$ GPa  (solid circles).
This indicates that the $k$-dependence is small in all the calculated systems, similar to the case of the diamond system presented in the main text.

 \appendix{ }  \section{Z-method }
The result of the Z-method is derived under the basic assumption $U_{\rm solid}(T_{\rm u})=U_{\rm liquid}(T_{\rm m})$, 
where $U_{\rm solid}(T)$ and $U_{\rm liquid}(T)$ are the internal energies of the system in the solid and liquid states, respectively.
Moreover,  two approximate relations, $C_V(T_{\rm m}) \simeq 3k_{\rm B}$ and $\Delta S\simeq k_{\rm B} \ln2$,  are used\cite{Z-method}, where $C_V (T_{\rm m})$
 is the specific heat at a constant volume in the solid state at  $T_{\rm m}$, and $\Delta S $ is the entropy change between the liquid and solid states\cite{Stishov1974}.
These lead to $T_{\rm u}/T_{\rm m} =( 1+\Delta S/C_V(T_{\rm m}) )\simeq1.23$\cite{Belonoshko2006}.

The first assumption is based on the results of MD simulations and seems plausible. 
To confirm the validity of the second relations, we compared them with experimental results, where the difference between $C_V(T_{\rm u}) $ and $C_P(T_{\rm u}) $ was neglected.
Table \ref{tableA1}  gives these experimental values with   melting points  $T^{\rm ex}_{\rm m}$ for several elements and compounds.\cite{NIST}
For reference, our calculated results of  $T_{\rm m}$ were also added.
\begin{table}[h]
\caption{Experimental data for $C_{\rm p}$, $\Delta S $, $\Delta S /C_{\rm p}$, and $T^{\rm ex}_{\rm m}$ together with  calculated $T_{\rm m}$ }
\begin{tabular}{l|ccccccc} \hline
                                                      &     Na         &   Ca        &    Al      &  NaCl    &  B$_4$C  &  Be$_2$C   \\          \hline                                                 
            $C_{\rm p}/k_{\rm B}$ &    3.80   &  5.20    &  3.96    & 4.05     &  3.79       &   3.69        \\ 
             $\Delta S  /k_{\rm B}$  &    0.844 &  0.921   &  1.38    & 1.58      &  0.917     &  1.26    \\ 
            $\Delta S  /C_{\rm p}$  &    0.222 &  0.177    &  0.242  & 0.389  &  0.242   &  0.341    \\   
   $T^{\rm ex}_{\rm m}$\cite{NIST} [K]   &    371     & 1115       &  933       & 1074    &  2743   &  2400        \\  \hline
   $T_{\rm m}$ [10$^3$K]        &  0.36    & 1.2       &  0.94       & 0.90      & 3.0     &  2.7     \\
    size $(N)$                                   &   16      & 32         &  4           &  8         &  45        &   12        \\     \hline
\end{tabular}
\label{tableA1}
\end{table}

 Although the  experimental values of $C_{\rm p}/k_{\rm B}$  and $\Delta S/k_{\rm B}$\cite{NIST} are larger than the theoretically expected values,  the values of $1+\Delta S  /C_{\rm p}$ are relatively close to 1.23.
The above results suggest that the equation $T_{\rm m}=T_{\rm u}/1.23$ is a reasonable approximation even for compounds.

To consider the validity of $T_{\rm m}$, we present some additional results on $T_{\rm m}$ below.
It is  431 K for Na with $N=54$,   1504 K for Ca with $N=$4,  and 585 K for Al with $N=32$.
The results show that the size dependence of $T_{\rm m}$ is not very consistent, and  $T_{\rm m}$ for larger systems does not necessarily give better values.
Although it may be difficult to precisely estimate  $T_{\rm m}$ by this method, it is possible to obtain  the approximate  $T_{\rm m}$ of complex systems such as compounds with a small amount of calculation.
%
%

 \appendix{ }  \section{ Estimation of $T_{\rm c}$}

To estimate the transition temperature $T_{\rm c}$ of  superconductivity,  we calculated  the Eliashberg electron phonon spectral function $\alpha^2 F(\omega)$ within the framework of density functional perturbation theory(DFPT)\cite{Baroni2001-DFPT} in  the QE package, where $\omega$ is the phonon frequency.
Here, $\alpha^2 F(\omega)$ is defined by
$$ \alpha^2 F(\omega)=\frac{1}{N(\varepsilon_{\rm F})}\sum_{mn}\sum_{\bf q\nu}\delta(\omega- \omega_{{\bf q} v})\sum_{\bf k}\vert g^{\bf q \nu,mn}_{\bf k+\bf q, \bf k}\vert^2$$
$$\times \delta(\varepsilon_{\bf k+\bf q, m} - \varepsilon_{\rm F}) \delta(\varepsilon_{\bf k, n} - \varepsilon_{\rm F}),$$
where  $g^{\bf q \nu,mn}_{\bf k+\bf q, \bf k}$ is the matrix element of the electron-phonon coupling  and $N(\varepsilon_{\rm F})$  is the density of states at the Fermi energy $\varepsilon_{\rm F}$  per both spins.
Within DFPT,  the electron-phonon matrix elements are obtained from the first-order derivative of the self-consistent Kohn-Sham potential. 
The summation over the Fermi surface is performed using the tetrahedron method.\cite{QE,Kawamura2014}

The logarithmic average of phonon frequency $\omega_{\rm  log}$ is written as
$$\omega_{\rm  log}=\exp\{\frac{2}{\lambda} \sum_{\bf q\nu  {\bf k} \rm mn}\frac{\log( \omega_{{\bf q} \nu})}{N(\varepsilon_{\rm F})\omega_{\bf q \nu}} \vert g^{\bf q \nu, \rm mn}_{\bf k+\bf q, \bf k}\vert^2 $$
$$ \times \delta(\varepsilon_{\bf k+\bf q, \rm m} - \varepsilon_{\rm F}) \delta(\varepsilon_{\bf k, \rm n} - \varepsilon_{\rm F})\}.$$
The  electron-phonon coupling constant $\lambda$ is given by
   $$\lambda =2\int_0^\infty \frac{\alpha^2 F(\omega)}{\omega} {\rm d}\omega.$$
The transition temperature $T_{\rm c}$ of  superconductivity was  calculated using  the Allen and Dynes formulation\cite{Allen1975}, 
$T_{\rm  c}=f_1 f_2 \frac{\omega_{\rm  log}}{1.2}\exp(-\frac{1.04(1+\lambda)}{\lambda-\mu^*(1+0.62\lambda) }),$
where, the  factors $f_1$ and $f_2$ are decided by the $\lambda$, $\mu^*$, $\omega_{\rm  log}$ and  mean square frequency
 $\langle \omega^2 \rangle$.
They are given by  
$$f_1=[1+(\lambda/\{2.46(1+3.8\mu^*)\}^{3/2}]^{1/3}$$
$$f_2=1+\frac{(\langle \omega^2 \rangle^{1/2}/\omega_{\rm  log}-1)\lambda^2}{\lambda^2+[1.82(1+6.3\mu^*)(\langle \omega^2 \rangle^{1/2}/\omega_{\rm  log})]^2 },$$
where  $\langle \omega^2 \rangle=\frac{2}{\lambda} \int_0^\infty \alpha^2 F(\omega)\omega  {\rm d}\omega $.
The screened Coulomb potential parameter  $\mu^*$  is  set to $0.1$ as a typical value.

In our calculation, $\bm {k}$ mesh ($\bm {q}$ mesh)  is chosen to be $8 \times 8  \times 8$ ($4 \times 4  \times 4$).
 In the optimization of structures, we used  convergence thresholds  as  $3 \times 10^{-6}$ a.u. in total energy and  $3 \times 10^{-5}$ a.u. in force.
To verify the validity of our estimation, we calculated $T_{\rm c}$ of YH$_6$ at $P=166$ GPa. We  found that  $T_{\rm c} \simeq 259$ K, while the experimental result at the same pressure is 224 K.\cite{Troyan-2019-YH6,Kong-2019-YH6}
For LaH$_{10}$, $T_{\rm c}$ is estimated to be 212 K at $P=220$ GPa, but it is 245 K in the experimental results.\cite{Drozdov-LaH10,Somayazulu-2019}.
These results suggest that the above formulation successfully derives $T_{\rm c}$ for phonon high-Tc superconductors.

As an example,   the result of   FC$_{10}$ at $P=$0 GPa is shown.
In this case,  we obtained  $\omega_{\rm  log}  \simeq 697$ K, $\lambda \simeq 1.78$, and $T_{\rm c} \simeq 106$ K. 
We confirmed that the results of  $T_{\rm c}$  are almost the same even if the meshes are different.
For example, if  $\bm {k}$  and $\bm {q}$  meshes were chosen to be $6 \times 6 \times 6$ and $3 \times 3 \times 3$, respectively, we obtained a result of $T_{\rm c} \simeq 112$ K.
For $12 \times 12  \times 12$ and $6 \times 6  \times 6$,  we obtained $T_{\rm c} \simeq 104$ K. 
%

\begin{figure}[thb]
\begin{center}
\includegraphics[width=0.8 \linewidth]{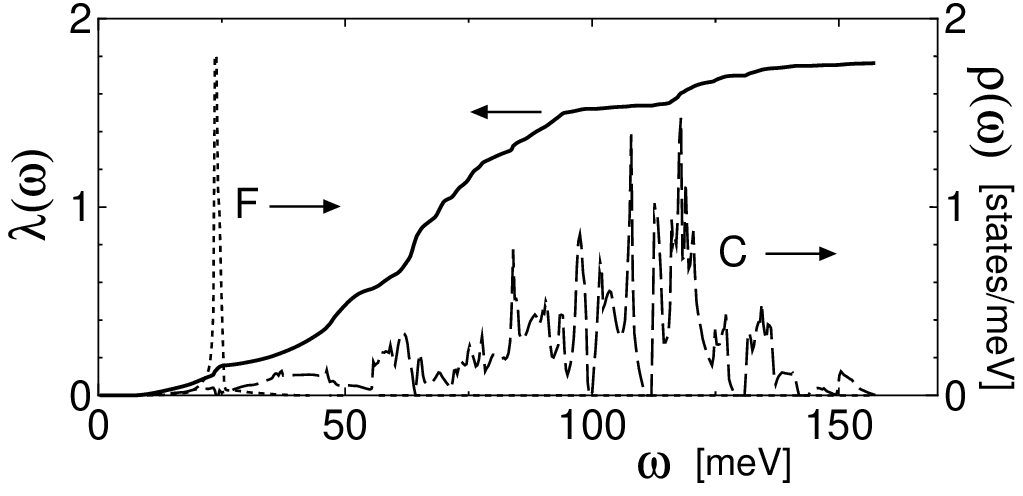}
\end{center}
\caption{ Electron-phonon coupling constant $\lambda(\omega)$  and  phonon DOS $\rho(\omega)$ for  FC$_{10}$ at $P=0$ GPa. Here,  the solid line represents $\lambda(\omega)$, the dotted line represents the contributions of fluorine atoms to $\rho(\omega)$, and the dashed line represents the contributions of carbon atoms  to $\rho(\omega)$.  
 }
\label{fig-c1}
\end{figure}

  Using $\alpha^2 F(\omega)$,  we defined  $\lambda (\omega)$ as 
$\lambda (\omega)=2\int_0^\omega \alpha^2 F(\omega')/\omega' {\rm d}\omega'.$
This allows us to see the contribution of each atom to  $\lambda$.
Figure \ref{fig-c1}   shows  $\lambda (\omega)$ as a function of $\omega$ (solid line).
 Figure \ref{fig-c1} also shows the phonon DOS decomposed  into the fluorine and carbon atoms, where the dotted line represents  the contributions of fluorine atoms, and the dashed line represents the contributions of carbon atoms.  
It indicates that the phonon spectrum is clearly divided into two parts: components of   F  and C atoms.
Furthermore, the contribution of phonons to $\lambda (\omega)$  is also   divided into two parts, and the F atom component seems to be about 1/8 of the C atom component.
This result indicates that it is mainly the carbon atoms that contribute to the superconductivity of FC$_{10}$.
Similar results have already  been obtained for  $X$C$_6$.\cite{Sano2022}

%

%
%
 \appendix{ }  \section{Free Energy }
We considered the Helmholtz   free energy  $F(T)$ per atom  within the harmonic approximation for phonons.
In  the harmonic approximation, the volume of the system does not depend on $T$.
Thus, we  neglected the volume change of the system   due to temperature hereafter.

Using the free energy formulation for a harmonic oscillator, we obtained
$$  F(T)= \int [ \frac{h \omega}{2}  + k_{\rm B}T \ln(1-\exp(- \frac{h \omega}{k_{\rm B}T}) ] \rho_{\rm ph}(\omega)d\omega, $$
where  $\rho_{\rm ph}(\omega)$ is the phonon DOS per atom, which is calculated by  QE.
The internal energy $U(T)$ was  also  obtained by a similar formulation as
$$  U(T)= \int  \frac{h \omega}{2}[1  +  \frac{1}{\exp(\frac{h \omega}{k_{\rm B}T} )-1}] \rho_{\rm ph}(\omega)d\omega. $$
Here, entropy $S(T)$ can be easily obtained by $S(T)=(U(T)-F(T))/T$.

 To compare the experimental results with the above results, we considered the  enthalpy of phonons, $H(T)$.
In our approximation, the $pV$ term in the enthalpy is independent of temperature.
Therefore, $H(T)$ as a function of $T$ is equivalent to  $U(T)$,  except  for a constant.
Since enthalpy is typically  used in  experiments\cite{NIST}, we used $H(T)$ instead of $U(T)$ to avoid confusion.

\begin{figure}[thb]
\begin{center}
\includegraphics[width=0.9 \linewidth]{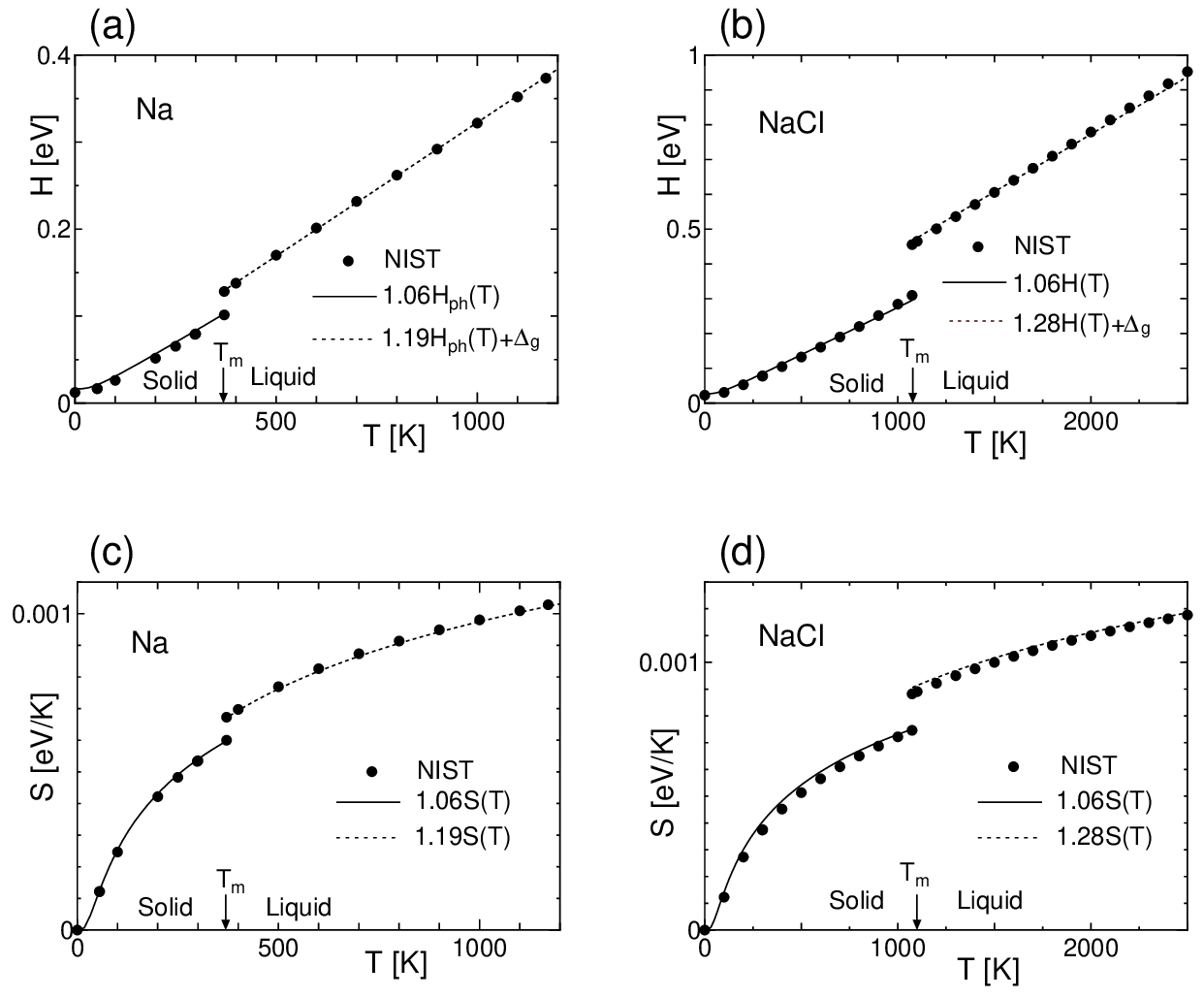}
\end{center}
\caption{Theoretical and  experimental results of enthalpy as a function of $T$  for (a) Na and (b)  NaCl,  and  those of entropy  for (c) Na and  (d) NaCl, where the experimental results are taken from NIST\cite{NIST}.  Here, the experimental values of enthalpy were shifted by a constant to match the  theoretical value of $H(T)$ at $T=0$. 
 }
\label{figB1}
\end{figure}

As shown in Fig. \ref{figB1}(a), we found that the experimental result of the  enthalpy $H^{\rm ex}(T)$ for sodium\cite{NIST} is well fitted by
$ H^{\rm ex}(T) \simeq c_{\rm solid}H(T)$   for $T \le T_{\rm m}$ and $H^{\rm ex}(T) \simeq c_{\rm liquid}H(T)+\Delta_g $   for $T > T_{\rm m}$, where $c_{\rm solid}$ is a constant, 1.06, and $c_{\rm liquid}$ is 1.19.  
Here, $\Delta_g$ is the value  to reproduce the gap in $H^{\rm ex}(T)$ at $T_{\rm m}$, which will be given later.
For the case of NaCl\cite{NIST}, we obtained $c_{\rm solid}=1.06$  and $c_{\rm liquid}=1.28$, as shown in Fig. \ref{figB1}(b). 
The $c_{\rm solid}$ values  are the same for Na and NaCl, but the $c_{\rm liquid}$  values are slightly different.

As shown  in Figs. \ref{figB1}(c) and \ref{figB1}(d), we  found that  similar relationships hold for entropies.
That is,  $ S^{\rm ex}(T) \simeq c_{\rm solid}S(T)$   for $T < T_{\rm m}$ and $S^{\rm ex}(T) \simeq c_{\rm liquid}S(T)$ for $T \geq T_{\rm m}$.
These relations lead to $\tilde{F}(T) \simeq c_{\rm solid}F(T)$   for $T < T_{\rm m}$   and $\tilde{F}(T) \simeq c_{\rm  liquid}F(T)+\Delta_g$ for $T \geq T_{\rm m}$, where  $\tilde{F}(T)$ is the phenomenologically corrected
 free energy that is  fitted to closely reproduce the experimental results, and $\Delta_g=(c_{\rm solid}-c_{\rm liquid})F(T_{\rm m})$.

The above phenomenological relations give approximate free energies of the solid and liquid states which closely reproduce the experimental results.
The results shown in Fig. \ref{figB1} suggest that this approximation can be widely applied to a variety of materials.
In fact, we confirmed that this method can also be applied to Al and B$_4$C\cite{NIST}.
We found that $c_{\rm solid}=1.09$ and $c_{\rm liquid}=1.3$ in the former case. In the latter case, $c_{\rm solid}=1.03$ and $c_{\rm liquid}=1.13$.

%
 
\begin{figure}[hb]
\begin{center}
\includegraphics[width=0.8 \linewidth]{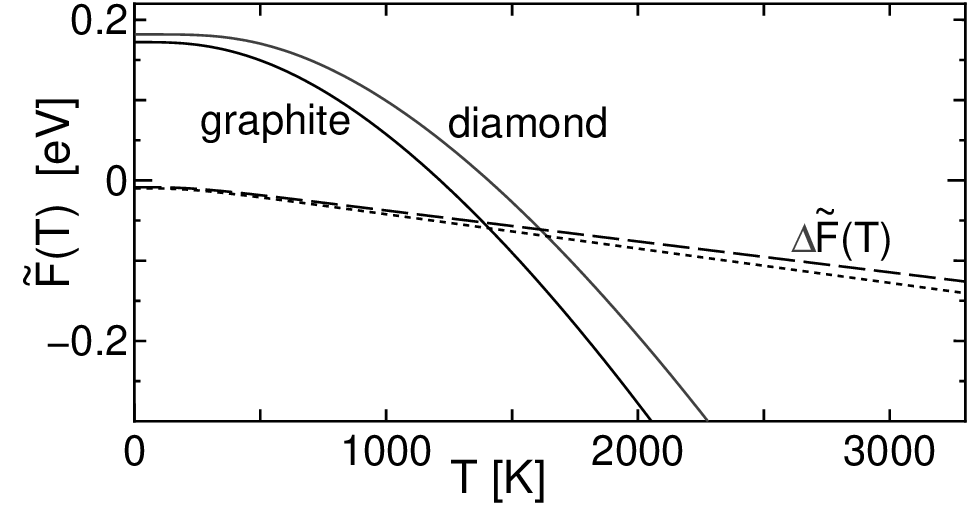}
\end{center}
\caption{ Phenomenological  free energies $\tilde{F}(T)$   of diamond and graphite (solid lines) at $P=8$ GPa, together with the formation   free energies $\Delta \tilde{F}(T)$ of  graphite at $P=8$  (dotted line) and $P=15$ GPa (dashed line).
 }
\label{figB2}
\end{figure}

To test the above analysis, we tried to find the phase boundary between graphite and diamond.
When $T=0$,  the enthalpy of diamond  is lower than that of  graphite and diamond is stable for $P \simj 2$ GPa\cite{Bundy1996,Ghiringhelli2005}. 
As the temperature increases, the free energy of graphite decreases more quickly than that of diamond, and graphite becomes  stable.
The boundary is given by  $\Delta \tilde{F}(T) + \Delta H = 0$, where $\Delta \tilde{F}(T)$ and  
 $\Delta H$ are the formation free energy and the formation  enthalpy of graphite, respectively.

Figure \ref{figB2} shows  the $\tilde{F}(T)$  of diamond and graphite and the $\Delta \tilde{F}(T)$ of graphite,  where $c_{\rm solid}$ is set to 1.06 for both  diamond and graphite.
Here, we confirmed that  the $\tilde{F}(T)$ of  graphite  almost reproduces the experimental result \cite{NIST}.
In the figure, the solid lines represent the $\tilde{F}(T)$  of diamond and graphite at $P=8$ GPa.
$\Delta \tilde{F}(T)$ is shown as a dotted line for $P=8$ GPa and as a dashed line for $P=15$ GPa.
The two lines are close to each other and look nearly identical.
The result indicates that the $P$-dependence of $\Delta \tilde{F}(T)$ is very small.
On the other hand, the $P$-dependence of $\Delta H$ is not so small and dominates the conditions that determine the phase boundary.

The  $\Delta H$  values are $-0.129$, $-0.0283$, 0.0318, 0.0514, and 	0.1293 for $P=0$, 5, 8, 10, and 15 GPa, respectively. 
Using the results above, we found that the boundary temperatures are approximately 0, 750, 1300, and 3400 K for $P=$ 6.9, 8, 10, and 15 GPa, respectively.
This boundary appears to be shifted by about 5 GPa toward higher pressure compared with  other  theoretical\cite{Ghiringhelli2005}   and experimental\cite{Bundy1996} results.
Apart from this, the result seems to agree with those of other works.
%
%
\begin{figure}[hb]
\begin{center}
\includegraphics[width=0.8 \linewidth]{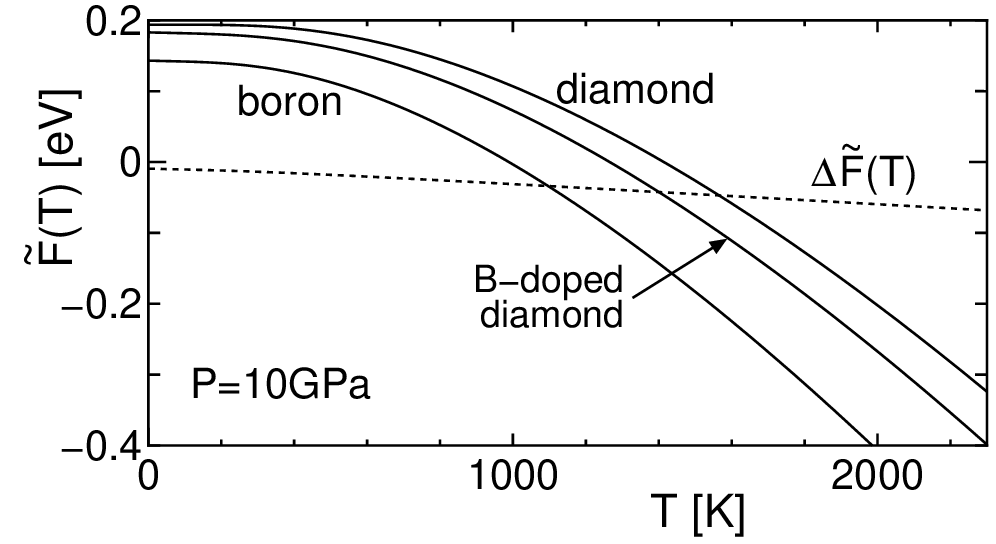}
\end{center}
\caption{Phenomenological  free energies $\tilde{F}(T)$   of diamond, boron, and boron-doped  diamond  at $P=10$ GPa (solid lines).
The formation  free energy $\Delta \tilde{F}(T)$ of   boron-doped  diamond  is represented by a dotted line.
 }
\label{figB3}
\end{figure}

The application to  the boron-doped diamond system is also interesting.
It is  synthesized at high pressure ($ \sim 10$ GPa) and temperature ( $ \sim 3000$ K) and shows superconductivity with $T_{\rm c} \sim 4$ K\cite{Ekimov2004}.  Here, the boron concentration $\rho_{\rm b}$ is about 3 percent.
 Figure \ref{figB3}  shows  the $\tilde{F}(T)$ of  diamond, boron-doped diamond, and boron at $P=10$ GPa.
For boron-doped diamond, we used the N = 32($4\times 4 \times 2$)  system, in which one atom is replaced by boron.

In a real system, the effect of entropy  becomes important\cite{Manzoor2018} because the boron atoms are randomly distributed in the system.
We evaluated the additional entropy by treating the boron-doped system as a type of binary alloy\cite{Lattice-gas}.
In this case, the entropy  is expressed as $ -k_{\rm B} (\rho_{\rm b} \ln \rho_{\rm b} + (1-\rho_{\rm b}) \ln (1-\rho_{\rm b}) )$.
For  $\rho_{\rm b}=0.03$,  the value is  $\simeq 0.12 \times 10^{-4}$ eV/K, and  we add it to the $\tilde{F}(T)$ of the boron-doped diamond.

In this figure, the dotted line indicates the formation free energy  $\Delta \tilde{F}(T)$ of the boron-doped  diamond.
Here, to obtain  $\Delta \tilde{F}(T)$, we assumed that the crystal structure of boron is $R\bar{3}m$.\cite{Material-project,Parakhonskiy2011}
It decreases with increasing temperature, and  $\Delta \tilde{F}(T)+ \Delta H$ becomes negative at $T \simj 1200$ K. 
Here,   $\Delta H$ is  the formation enthalpy of the  boron-doped  diamond, and its value  is about 0.038 eV.
This result means that the  boron-doped  diamond is stable at $T \simj 1200$ K\cite{B-dope}.
It  does not contradict the synthesis conditions of the boron-doped diamond\cite{Ekimov2004}.
%

As mentioned above, our evaluation of the free energy  is expected to be useful in finding conditions for material synthesis.
To analyze the free energy of NaC$_6$, we used $c_{\rm solid}=1.06$ and $c_{\rm liquid}=1.13$ as the phenomenological parameters for sodium at $P=300$ GPa.
For simplicity, we also used  $c_{\rm solid}=1.06$   for  diamond and NaC$_6$. 
The calculated  results are given in the main text.
%



\begin{thebibliography}{99}
\bibitem{NaC6}
To precisely discuss the synthesis conditions, it may be necessary to improve the accuracy of $\Delta H$. 
Additionally,  the presence of other compounds such as NaC may need to be considered.
\bibitem{Z-method}
If the simulations can be performed in sufficient detail, it is possible to determine $T_{\rm m}$ without using these two approximate relations.
\bibitem{B-dope}
 We confirmed that these results are almost unchanged even if the existence of B$_4$C is taken into account.
\end{thebibliography}
\end{document}